\documentclass[
aps,
pra,
twocolumn,
superscriptaddress,
longbibliography,
amsmath,
amssymb,
floatfix
]{revtex4-2}

\usepackage{graphicx}
\usepackage{physics}
\usepackage{bm}
\usepackage{amsmath,amssymb}
\usepackage{amsthm}
\usepackage{hyperref}
\usepackage{booktabs}

\hypersetup{
colorlinks=true,
linkcolor=blue,
citecolor=blue,
urlcolor=blue
}

\begin{document}

\title{Efficient Quantum Simulation of Linearized Vlasov--Poisson Dynamics Using Trotter and THRIFT Hamiltonian Simulation Methods}

\author{Kartick Paul}
\email{kartickpaul6290@gmail.com}

\author{Rahul V}
\email{ph22d002@iittp.ac.in}

\author{S. Aravinda}
\email{aravinda@iittp.ac.in}

\author{Reetesh K. Gangwar}
\email{reetesh@iittp.ac.in}
\affiliation{Department of Physics, Indian Institute of Technology Tirupati, Tirupati 517619, Andhra Pradesh, India}

\begin{abstract}
The Vlasov--Poisson system provides the fundamental kinetic description of plasma and plays a central role in understanding collective phenomena such as Landau damping and wave--particle interactions. Efficient numerical simulation of these dynamics remains challenging because of the high dimensionality of phase space. In this work, we studied magnetized and non-magnetized plasma using a quantum simulation framework for the linearized Vlasov-Poisson equation by reformulating the discretized system as a Hermitian Hamiltonian suitable for gate-based quantum computation. The time evolution is implemented using first, second and fourth-order Trotter--Suzuki product formulas and the recently proposed Time-Resolved Interaction Framework (THRIFT). The performance of the different simulation methods is systematically evaluated through electric field evolution, state fidelity, convergence behavior, energy conservation, entanglement entropy and quantum resource requirements, including circuit depth and two-qubit gate complexity, for both magnetized and non-magnetized plasma models. To further reduce finite time step errors without increasing circuit depth, Richardson extrapolation is incorporated as a error-mitigation technique. The results provide a comprehensive comparison of Trotter and THRIFT approaches and establish practical guidelines for accurate and resource-efficient quantum simulation of plasma dynamics on gate-based quantum computers.
\end{abstract}
\maketitle
\section{Introduction}
\noindent
Quantum computing has emerged as a fundamentally different paradigm for information processing, using quantum superposition, interference and entanglement to perform computational tasks in ways that are not directly available on conventional computers \cite{dalzell2025quantum,Huang2026Vast}. Quantum algorithms can provide substantial speedups for important computational problems, demonstrating the potential of quantum computing. Shor's algorithm can factor integers in polynomial time, while Grover's algorithm provides a quadratic speedup for unstructured search, reducing the query complexity from $\mathcal{O}(N)$ to $\mathcal{O}(\sqrt{N})$ \cite{Shor1994,Shor1997,Grover1996,Boyer1998}. 
These landmark results established that quantum algorithms can exploit the structure of computational problems to outperform known classical approaches and motivated the development of quantum algorithms beyond cryptography and search.
In particular, quantum simulation was identified as a natural application of quantum computers because physical systems are themselves governed by quantum and linear-algebraic principles \cite{Feynman1982,Lloyd1996}. A quantum register of $n$ qubits spans a Hilbert space of dimension $2^n$, allowing quantum states to encode high-dimensional vectors through their amplitudes. 
However, a potential computational advantage also depends on the costs of state preparation, quantum evolution, error control and measurement.
This capability has motivated the development of quantum algorithms for large-scale linear algebra, differential equations, partial differential equations and dynamical-system simulation \cite{Nielsen2010,Preskill2018,Harrow2009,Berry2014,Berry2017,Childs2020,Childs2021,Krovi2023improvedquantum,Costa2019wave,PhysRevResearch.7.023262}.

Differential equations are important problems in scientific computing and many quantum algorithms have been developed to solve them. Many physical systems are described by ordinary or partial differential equations, whose numerical discretization leads to large systems of coupled linear equations or linear operator evolution. Quantum algorithms for linear differential equations have been proposed using Hamiltonian simulation and quantum linear-algebraic techniques \cite{Berry2014,Berry2017,Krovi2023improvedquantum}. Quantum spectral and Hamiltonian-simulation methods provide routes to solving differential equations by representing solution spaces and differential operators on a quantum computer \cite{Childs2020,Costa2019wave,Jin2022quantumdifference}. 
Recent work has further demonstrated quantum algorithms for partial differential equations using Fourier-space constructions, quantum singular-value transformation, block encoding and higher-order discretization techniques \cite{Lubasch2025PDEFourier,Guseynov2025BlockEncoding,Arseniev2025HighOrderPDE,Sato2025NonconservativePDE}.
These developments are particularly relevant for high-dimensional transport and field equations, where the choice of discretization, operator representation and quantum evolution method directly affects the required quantum resources.
Quantum algorithms can also solve some partial differential equations with high numerical precision while maintaining favorable computational scaling \cite{Childs2021}.

One particularly important class of scientific applications of quantum computing is the simulation of dynamical systems governed by Hamiltonian evolution. Hamiltonian simulation provides a natural framework for implementing the time evolution of physical systems on quantum computers and has therefore become a central problem in quantum algorithms \cite{Lloyd1996,Childs2021}. A major challenge is to approximate the required time-evolution operator accurately using a finite sequence of elementary quantum gates while controlling the associated computational cost. Several approaches have consequently been developed to improve the accuracy and resource requirements of Hamiltonian simulation, including product-formula methods, quantum signal processing and qubitization \cite{Berry2015,Low2019,Gilyen2019,Childs2021,Fang2023timemarchingbased,Berry2024quantumalgorithm}. 
Recent studies have also explored explicit block-encoding constructions, application-oriented benchmarking and alternative randomized or hybrid strategies for Hamiltonian simulation \cite{Guseynov2025BlockEncoding,Granet2025AppQSim,Fan2025AdaptiveRandomCompiler,Gentinetta2025CorrectingTrotter}.
Among these approaches, product-formula methods are particularly attractive for gate-based quantum simulation because they provide a systematic decomposition of complex Hamiltonian dynamics into elementary unitary operations.

Quantum computation is especially interesting for plasma physics because kinetic plasma models describe the evolution of distribution functions in a phase space containing both spatial and velocity degrees of freedom. The Vlasov--Poisson equation governs the dynamics of a collisionless plasma, where the distribution function evolves under self-consistent electromagnetic forces \cite{Vlasov1968,Nicholson1983,Chen2016,Ameri2023linearVlasov,Ye2024QTN,Higuchi2023kinetic,Berntson2026weakly}. As the number of spatial and velocity dimensions increases, the computational cost required to accurately resolve the distribution function grows rapidly \cite{Ye2022VlasovMPS,Ye2024QTN,Higuchi2023kinetic}. Kinetic plasma dynamics therefore provide a useful setting in which to explore quantum methods for high-dimensional differential equations \cite{Engel2019,Costa2019wave,Ameri2023linearVlasov,Cappelli2024VlasovSchrodinger}. 
Quantum simulation of plasma waves has also recently been demonstrated on superconducting quantum hardware \cite{Sundar2026PlasmaChip}.
However, simply having fewer qubits does not ensure a computational advantage, since the costs of state preparation, time evolution and measurement must also be taken into account \cite{Engel2019,Ye2022VlasovMPS,Andress2025nonlinear,Berntson2026weakly,DiPiazza2025quantumplasma}. Consequently, quantum simulation of kinetic plasma dynamics requires not only a suitable quantum representation of the governing equations but also a careful analysis of accuracy, circuit complexity and error sources.

In recent years, quantum algorithms for kinetic plasma dynamics have started to emerge. Engel, Smith, and Parker framed the linearized Vlasov--Poisson system as a Hamiltonian simulation problem and examined its application to the electrostatic Landau-damping regime \cite{Engel2019,Jin2023SchrodingerizationPRA}. Their work demonstrated that linearized kinetic plasma dynamics can be represented within a quantum Hamiltonian framework and highlighted the dependence of computational requirements on measurement precision. More recently, Toyoizumi, Yamamoto, and Hoshino studied Hamiltonian simulation using quantum singular-value transformation (QSVT) for the one-dimensional linearized Vlasov--Poisson equation. They analyzed the simulation error and query complexity of QSVT-based Hamiltonian simulation and demonstrated the reproduction of linear Landau damping in the quantum framework \cite{Toyoizumi2024}. Quantum-inspired approaches have also been developed for the Vlasov--Poisson system, emphasizing the importance of exploiting the structure of kinetic plasma dynamics \cite{Ye2022,Jin2022quantumdifference,Jin2023linearrepresentations}. 
In parallel, recent quantum algorithms for advection--diffusion and drift--diffusion equations provide additional approaches for treating transport-dominated partial differential equations and enable explicit comparisons of Hamiltonian simulation, quantum Fourier-transform and other quantum algorithms \cite{Alipanah2025AdvectionDiffusion,Devereux2025DriftDiffusion}.
These studies demonstrate the growing interest in quantum and quantum-inspired approaches for the simulation of kinetic plasma dynamics.

Despite being introduced several decades ago \cite{Trotter1959,Suzuki1993}, product-formula methods based on the Trotter--Suzuki decomposition remain among the most widely used approaches for digital Hamiltonian simulation because of their conceptual simplicity, systematic convergence, and compatibility with gate-based quantum architectures \cite{Lloyd1996,Childs2021}. The approximation error can be reduced either by decreasing the time-step size or by employing higher-order product formulas. However, these improvements generally require additional applications of elementary evolution operators, thereby increasing the circuit depth and gate count \cite{Childs2021,Hu2024quantumcircuits}. On current noisy intermediate-scale quantum (NISQ) devices, increased circuit depth and, in particular, the number of two-qubit gates can substantially amplify the effects of gate errors and decoherence \cite{Preskill2018,An2025QDESolvers}. 
Recent work has shown that Trotter errors are closely connected to quantum resources such as entanglement and nonstabilizerness, while other studies have investigated ways of mitigating Trotter errors through extrapolation, unitary corrections and hybrid quantum-classical constructions \cite{Zhang2026TrotterResources,Watson2025TrotterMitigation,Zeng2025TrotterLCU,Gentinetta2025CorrectingTrotter}.
These developments reinforce the importance of evaluating Hamiltonian simulation methods not only by their formal error order, but also by their practical circuit-depth and gate-count requirements.
Consequently, achieving a favorable balance between simulation accuracy and quantum-resource requirements remains a central challenge for practical Hamiltonian simulation.

Recently developed time-dependent product-formula and interaction-picture methods have emerged as promising alternatives to conventional product-formula approaches, particularly for Hamiltonians containing well-separated energy scales \cite{LowWiebe2019,Rajput2022,Bosse2025}. By treating a dominant part of the Hamiltonian exactly and approximating the remaining interaction terms, these methods can exploit Hamiltonian structure to improve simulation accuracy and resource scaling. 
In particular, Bosse \textit{et al.} introduced the Trotter Heuristic Resource Improved Formulas for Time-dynamics (THRIFT) framework, which combines an interaction-picture formulation with product formulas and is designed to exploit separated energy scales and hardware-efficient implementations \cite{Bosse2025}.
Although THRIFT has demonstrated favorable accuracy and resource scaling for selected Hamiltonian models, 
its performance for kinetic plasma Hamiltonians has not, to our knowledge, been systematically investigated.
Moreover, recent work on time-dependent Trotter-error bounds provides a natural theoretical framework for extending product-formula methods to Hamiltonians with explicit time dependence \cite{Mizuta2026TimeDependentTrotter}.
In particular, for the linearized Vlasov--Poisson system, the trade-off between simulation accuracy and quantum-resource requirements, including two-qubit gate count, circuit depth, and CNOT depth, remains insufficiently explored. This motivates a systematic comparison of THRIFT and conventional Trotter--Suzuki methods in terms of accuracy, error scaling, and quantum circuit resources.
In addition to algorithmic error, practical quantum simulations must account for hardware noise and the propagation of errors through digital Hamiltonian evolution. Recent studies have investigated error dilution in digital Hamiltonian simulation, application oriented benchmarking of Hamiltonian simulation tasks and adaptive randomized compilation as alternative strategies for improving the practical performance of quantum circuits \cite{Granet2025DilutionError,Granet2025AppQSim,Fan2025AdaptiveRandomCompiler}.

In this work, we investigate a complete quantum simulation framework for the linearized Vlasov--Poisson system based on a Hermitian representation of the Hamiltonian. We study implementations of traditional Trotter--Suzuki product formulas and the THRIFT algorithm at first, second, and fourth order for both magnetized and non-magnetized plasma configurations. We study the entanglement entropy and quantum-resource metrics, including circuit depth, two-qubit gate count and CNOT depth to assess the computational cost and error behavior of the simulations. We employ Richardson extrapolation as a numerical error-mitigation technique to further reduce finite-time-step errors without increasing the depth of an individual quantum circuit. \cite{Watson2025TrotterMitigation}.We also analyze state fidelity, convergence behavior, and energy conservation. 
Together, these quantities allow us to characterize the accuracy--resource trade-off of the different Hamiltonian-simulation strategies at the level of both physical observables and quantum-circuit complexity.
Our results provide a comprehensive assessment of the trade-off between simulation accuracy and quantum resources and establish practical guidelines for accurate and resource-efficient quantum simulation of kinetic plasma dynamics on gate-based quantum computers.
\section{Hamiltonian Formulation from the Vlasov--Poisson System}
\label{1}
The kinetic evolution of a collisionless plasma is described by the phase-space distribution function $f_s(\mathbf{r},\mathbf{v},t)$ which specifies the density of particles of species $s$ at position $\mathbf{r}$ with velocity $\mathbf{v}$ and time $t$. Because electromagnetic interactions are long-range, plasma dynamics is inherently collective and is naturally described in the framework of kinetic theory.  The evolution of the distribution function, together with the self-consistent electric field $\mathbf{E}=(E_x,E_y,E_z)$ is governed by the non relativistic Vlasov--Poisson system\cite{Krall1973,Nicholson1983,Chen2016,Ye2022},
\begin{equation} \label{eq:1}
\frac{\partial f_s}{\partial t}
+\mathbf{v}\cdot \boldsymbol{\nabla}_{\mathbf r}f_s
+\frac{q_s}{m_s}
\left(\mathbf E+\mathbf v\times\mathbf B\right)
\cdot \boldsymbol{\nabla}_{\mathbf v}f_s
=0.
\end{equation}
\begin{equation}
\frac{\partial \mathbf{E}}{\partial t}
= -\frac{q}{\epsilon_0} \int \mathbf{v} \, f(\mathbf{r}, \mathbf{v}, t)\, d\mathbf{v}
\end{equation}
Here, $q_s$ and $m_s$ denote the charge and mass of species $s$,
respectively. The vectors $\mathbf{E}$ and $\mathbf{B}$ denote the self consistent electric field and prescribed magnetic field, respectively. The operators $\boldsymbol{\nabla}_{\mathbf r}$ and
$\boldsymbol{\nabla}_{\mathbf v}$ denote the gradients with respect to configuration space and velocity space, respectively. Throughout this work, only the electron dynamics are considered, while the ions are treated as an immobile and uniform neutralizing background.

To investigate the plasma response in the vicinity of an equilibrium state, the electron distribution function is decomposed into an equilibrium component and a small perturbation as \cite{Nicholson1983,Chen2016,Engel2019,Ameri2023}
\begin{equation*}
\begin{aligned}
&f_s(\mathbf{r},\mathbf{v},t)
=
f_0(\mathbf{v})
+
f_1(\mathbf{r},\mathbf{v},t) \\[2mm]
&\mathbf{E}(\mathbf{r},t)
=
\mathbf{E}_1(\mathbf{r},t)
\hspace{0.5cm}\text{and}\hspace{0.5cm}
\mathbf{B}(\mathbf{r},t)
=
\mathbf{B}_0 .
\end{aligned}
\end{equation*}
Here, $f_0(\mathbf{v})$ denotes the equilibrium Maxwellian electron distribution, while $f_1(\mathbf{r},\mathbf{v},t)$ represents a small perturbation about the equilibrium state. For the homogeneous and quasi-neutral equilibrium considered here, the equilibrium electric field is assumed to vanish, $\mathbf{E}_0=0$. Hence the total electric field is given solely by the perturbation field $\mathbf{E}_1$. The plasma is embedded in a uniform and time-independent external magnetic field directed along the $z$ axis, $
\mathbf{B}_0=B_0\hat{\mathbf{z}},$ where $B_0$ is constant. The corresponding signed electron cyclotron frequency is defined as
$
\Omega_e=\frac{q_s B_0}{m_s},
$
where $q_s=-e$ and $m_s$ are the electron charge and mass, respectively. This frequency characterizes the gyromotion of electrons about the magnetic-field lines.

To facilitate the numerical treatment, the governing equations are first transformed into Fourier space with respect to the spatial coordinates. The resulting equations are then written in dimensionless form by introducing characteristic plasma scales and the corresponding normalized variables. The dimensionless variables are defined as \cite{Engel2019,Ameri2023}
\begin{equation} \begin{aligned} \hat{\mathbf{k}}&=\lambda_{De}\mathbf{k},\qquad \hat{t}=\omega_{pe}t,\qquad \hat{\mathbf{v}}=\frac{\mathbf{v}}{\lambda_{De}\omega_{pe}},\\ \hat{f}&=\frac{(\lambda_{De}\omega_{pe})^{3}}{n_e}f,\qquad \hat{\mathbf{E}}=\frac{e\lambda_{De}}{k_BT_e}\mathbf{E},\qquad \hat{\mathbf{B}}=\frac{e\lambda_{De}}{k_BT_e}\mathbf{B}. \end{aligned} \end{equation}
Here, $\mathbf{k}=(k_x,k_y,k_z)$ denotes the wave vector in Fourier space, $T_e$ is the temperature, $\omega_{pe}$ is the plasma frequency, $n_e$ is the unperturbed plasma density and $\lambda_{De}$ is the Debye length. For notational simplicity, the hats denoting dimensionless quantities are omitted in the following.
The velocity space is then discretized on a finite grid, with $\mathbf{v_l}=(v_{xl},v_{yl},v_{zl})$ denoting the $l$-th velocity-space grid point, where $l=1,2,\ldots,N_v$.
 
To obtain a Hermitian representation of the linearized Vlasov-Poisson system suitable for quantum simulation, we introduce the transformed
variables
\begin{equation}
\label{TV}
F_l
=
i\sqrt{\frac{\Delta V_v}{f_0(\mathbf{v}_l)}}\,
f(\mathbf{v}_l,t),
\qquad
\mu_l
=
\frac{q}{m v_T^2}
\sqrt{\Delta V_v\,f_0(\mathbf{v}_l)}.
\end{equation}
where $\Delta V_v=\Delta v_x\,\Delta v_y\,\Delta v_z$ denotes the
velocity-space volume element for a uniform Cartesian mesh having grid spacings $\Delta v_x$, $\Delta v_y$ and $\Delta v_z$.
This transformation symmetrizes the coupling between the perturbed distribution function and the electric field, thereby allowing the
linearized Vlasov--Poisson system to be expressed in terms of a Hermitian Hamiltonian suitable for quantum simulation
\cite{Toyoizumi2024}. Using the transformation defined in
Eq.~\eqref{TV}, the linearized Vlasov--Poisson system can be written as
a coupled set of ordinary differential equations,
\begin{equation}
\begin{aligned}
\frac{dF_l}{dt}
=&
-i(\mathbf{k}\!\cdot\!\mathbf{v}_l)F_l
-i\sum_{a=x,y,z}\mu_l v_{la}E_a
\\
&-\Omega_c
\left(
v_{ly}\frac{\partial F_l}{\partial v_x}
-
v_{lx}\frac{\partial F_l}{\partial v_y}
\right),
\end{aligned}
\label{eq:4}
\end{equation}
and
\begin{equation}
\frac{dE_a}{dt}
=
-i\sum_l \mu_l v_{la}F_l,
\qquad a=x,y,z.
\label{eq:5}
\end{equation}

Using the velocity-space discretization introduced above, Eqs.~\eqref{eq:4} and~\eqref{eq:5} form a finite-dimensional system of coupled ordinary
differential equations.  We collect the velocity-space amplitudes and the
electric-field components into the normalized state vector
\begin{equation}
|\Psi(t)\rangle
=
\frac{1}{\eta}
\begin{pmatrix}
F_1(t)\\
F_2(t)\\
\vdots\\
F_{N_v}(t)\\
E_x(t)\\
E_y(t)\\
E_z(t)
\end{pmatrix},
\label{eq:state_vector}
\end{equation}
where $\eta$ is a normalization factor chosen such that
$\langle\Psi(t)|\Psi(t)\rangle=1.
$
The resulting equations of motion can then be written in the
Schr\"odinger-like form
\begin{equation}
\frac{d}{dt}|\Psi(t)\rangle
=
-iH|\Psi(t)\rangle,
\label{eq:schrodinger}
\end{equation}
where $H$ is the Hermitian Hamiltonian associated with the linearized
Vlasov-Poisson dynamics.

The Hamiltonian is naturally decomposed into three physically distinct
contributions,
\begin{equation}
H
=
H_{\mathrm{stream}}
+
H_{\mathrm{electric}}
+
H_{\mathrm{magnetic}}
\label{eq:H_decomposition}
\end{equation}
corresponding to the free streaming, electric-field coupling and magnetic gyromotion, respectively. The streaming and magnetic contributions are represented in the
velocity-space basis by
\begin{equation}
\begin{aligned}
\left(H_{\mathrm{stream}}\right)_{lm}
&=
(\mathbf{k}\!\cdot\!\mathbf{v}_l)\delta_{lm},
\\
\left(H_{\mathrm{magnetic}}\right)_{lm}
&=
i\Omega_c
\left(
v_y\frac{\partial}{\partial v_x}
-
v_x\frac{\partial}{\partial v_y}
\right)_{lm}.
\end{aligned}
\label{eq:H_stream_magnetic}
\end{equation}
where $\delta_{lm}$ is the Kronecker delta.
The electric field coupling is described by
\begin{equation}
C_{la}=\mu_l v_{la},
\qquad a=x,y,z.
\label{eq:C_matrix}
\end{equation}
Thus, the complete Hamiltonian takes the block form
\begin{equation}
\label{eq:Hamiltonian}
H=
\begin{pmatrix}
H_{\mathrm{stream}}+H_{\mathrm{magnetic}} & C\\
C^\dagger & 0
\end{pmatrix}.
\end{equation}
Here, $C$ couples the $N_v$ velocity-space amplitudes to the three
electric-field components. With the magnetic differentiation operator
discretized appropriately, the resulting Hamiltonian is Hermitian,
$H^\dagger=H$ which ensures unitary time evolution.

\section{Quantum Computational Simulation }
The Hermitian Hamiltonian formulation derived in the previous section provides the starting point for the digital quantum simulation of the linearized Vlasov-Poisson dynamics \cite{Nielsen2010}. After discretization of velocity space, the
system is represented by a finite-dimensional state vector
$|\Psi(t)\rangle$, whose time evolution is governed by
\begin{equation}
|\Psi(t)\rangle
=
U(t)|\Psi(0)\rangle
=
e^{-iHt}|\Psi(0)\rangle,
\end{equation}
\label{eq:quantum_evolution}
where $H$ is the Hermitian Hamiltonian defined in
Eq.~\eqref{eq:H_decomposition} and $U(t)=e^{-iHt}$ is the corresponding unitary time-evolution operator. The initial state is constructed from the prescribed perturbation of
the equilibrium distribution and the corresponding self-consistent electric-field amplitudes. The state is subsequently normalized to satisfy $\langle\Psi(0)|\Psi(0)\rangle=1$ before applying the time-evolution
operator. To simulate the Vlasov–Poisson system on a quantum computer, we need to efficiently implement the time-evolution operator $U(t)$ as a quantum  circuit.\cite{Feynman1982,Lloyd1996,LowWiebe2019,Childs2021}.

For realistic plasma Hamiltonians, the different physical contributions generally do not commute. So, the exact propagator $U(t)$ cannot, in general, be directly implemented as a sequence of elementary quantum gates. Approximate Hamiltonian simulation methods are therefore needed to implement the time evolution on a gate-based quantum computer using a sequence of quantum gates.\cite{Nielsen2010,JavadiAbhari2024}.

In this work, we investigate two product-formula approaches for the digital simulation of the linearized Vlasov-Poisson dynamics: the
conventional Trotter-Suzuki method
\cite{Trotter1959,Suzuki1993} and the recently proposed Trotter Heuristic Resource Improved Formulas for Time-dynamics (THRIFT)
framework \cite{LowWiebe2019,Bosse2025}. Both approaches are applied to the same Hermitian plasma Hamiltonian defined in Eq.~\eqref{eq:H_decomposition} which allows their performance to be assessed under identical physical and numerical conditions. This
provides a consistent framework for examining how the choice of product-formula construction affects the approximation error and the
quantum resources required to reproduce the plasma dynamics. We therefore evaluate both approaches for different product-formula
orders and numbers of time-evolution steps which provides a systematic comparison of their accuracy and computational cost.

To facilitate a systematic comparison, the Hamiltonian is partitioned as
\begin{equation}
H = H_0 + \alpha H_1,
\label{eq:H_partition}
\end{equation}
In the present formulation, $H_0$ corresponds
to the streaming contribution, while $H_1$
contains the remaining interaction
terms and $\alpha$ characterizes its relative
strength. This partition provides a common framework for analyzing the convergence and error scaling of both simulation approaches. For a total evolution time $t$ divided into $r$ product-formula steps, the leading-order global errors of the first, second and fourth-order
Trotter-Suzuki formulas scale as
\begin{equation}
\mathcal{O}\!\left(\frac{\alpha t^2}{r}\right),\qquad
\mathcal{O}\!\left(\frac{\alpha t^3}{r^2}\right),\qquad
\mathcal{O}\!\left(\frac{\alpha t^5}{r^4}\right),
\label{eq:trotter_scaling}
\end{equation}
respectively, under the perturbative scaling assumed in the
Hamiltonian decomposition \cite{Childs2021,Bosse2025}. The corresponding
derivation is provided in Appendix~\ref{app:trotter}.

In the THRIFT formulation, the dominant Hamiltonian $H_0$ is treated
exactly in the interaction picture, while the remaining interaction
Hamiltonian is approximated using product formulas
\cite{LowWiebe2019,Bosse2025}. Under the assumptions of the THRIFT error
analysis, the leading-order global errors for the first, second and fourth-order formulations scale as
\begin{equation}
\mathcal{O}\!\left(\frac{\alpha^2 t^2}{r}\right),\qquad
\mathcal{O}\!\left(\frac{\alpha^2 t^3}{r^2}\right),\qquad
\mathcal{O}\!\left(\frac{\alpha^2 t^5}{r^4}\right),
\label{eq:thrift_scaling}
\end{equation}
respectively \cite{Bosse2025}. The corresponding derivation is given in
Appendix~\ref{app:thrift}. Thus, in the perturbative regime
$\alpha\ll1$, the interaction-picture construction can suppress the
leading dependence of the error on the interaction strength from
$\mathcal{O}(\alpha)$ to $\mathcal{O}(\alpha^2)$. This suppression is
particularly advantageous when the Hamiltonian exhibits a clear
separation between the dominant and interaction energy scales.
\subsection{Quantum-State Encoding and Measurement}
After discretization and padding of the Hamiltonian dimension to the nearest power of two, the plasma state is encoded into an $n$-qubit quantum register $|\Psi(t)\rangle$ defined in Eq.~\eqref{eq:state_vector}, where
\begin{equation}
n=\left\lceil \log_2 N \right\rceil,
\end{equation}
with $N$ denoting the padded Hilbert-space dimension. The normalized initial state is expressed in the computational basis as
\begin{equation}
|\Psi(0)\rangle
=
\sum_{j=0}^{N-1}c_j(0)|j\rangle,
\qquad
\sum_{j=0}^{N-1}|c_j(0)|^2=1.
\end{equation}
The quantum circuit then implements the approximate unitary time-evolution operator obtained from either the Trotter-Suzuki or
THRIFT product-formula decomposition,
\begin{equation}
|\Psi(t)\rangle
=
U(t)|\Psi(0)\rangle
=
\sum_{j=0}^{N-1}c_j(t)|j\rangle.
\end{equation}

At each evolution time, measurements are performed in the computational basis. Repeating the circuit for
$N_{\mathrm{shots}}$ shots gives an estimator for the probability of
each basis state according to Born's rule,
\begin{equation}
P_j(t)
=
|\langle j|\Psi(t)\rangle|^2
=
|c_j(t)|^2,
\end{equation}
with
\begin{equation}
P_j(t)\simeq
\frac{N_j}{N_{\mathrm{shots}}},
\end{equation}
where $N_j$ is the number of measurement outcomes corresponding to
the state $|j\rangle$.
For the present encoding, the basis state associated with the electric-field degree of freedom is assigned to the index
$j=N_v$. Thus, its measurement probability is
\begin{equation}
P_E(t)=P_{j=N_v}(t)
\simeq
\frac{N_E}{N_{\mathrm{shots}}},
\end{equation}
where $N_E$ denotes the corresponding measurement count. Since the quantum state is normalized by the factor $\eta$ introduced in
Eq \eqref{eq:state_vector}, the physical electric-field amplitude is recovered as
\begin{equation}
|E(t)|^2=\eta^2 P_E(t).
\end{equation}
Consequently, repeated state preparation, time evolution and measurement at successive evolution times provide the temporal
evolution of the electric-field energy as shown in next section.
\section{Results and Discussion}

\subsection{Electric-Field Evolution in non-magnetized plasma}

We first investigate the digital quantum simulation of the linearized Vlasov-Poisson system in the absence of a magnetic field
($\Omega_c=0$). The temporal evolution of the self-consistent electric field provides a direct measure of the collective plasma response and the collisionless Landau damping caused by phase mixing \cite{Landau1946,Ryutov1999,Mendonca2023}.

The reference dynamics are obtained from the exact Hamiltonian propagator
\begin{equation}
    |\Psi(t)\rangle
    =
    e^{-iHt}|\Psi(0)\rangle ,
    \label{HP}
\end{equation}
The Hermitian Hamiltonian is diagonalized by eigenvalue decomposition as
\begin{equation}
H=V\Lambda V^\dagger,
\end{equation}
where $V$ is the unitary matrix whose columns are the eigenvectors of H and $\Lambda$ is the diagonal matrix containing the corresponding real eigenvalues.
The reference state can therefore be evaluated as
\begin{equation}
    |\Psi(t)\rangle
    =
    V e^{-i\Lambda t}V^\dagger|\Psi(0)\rangle .
    \label{eq:exact_evolution}
\end{equation}
The electric-field component is subsequently extracted from the evolved state. The corresponding approximate dynamics are obtained
using first and second-order Trotter--Suzuki and THRIFT product
formulas. The electric-field energy is defined as
\begin{equation}
    \mathcal{E}_E(t)=|E(t)|^2 ,
    \label{eq:electric_field_energy}
\end{equation}
and is used as the primary physical observable for comparison with the exact Hamiltonian evolution.

\begin{figure}[h]
    \centering
    \includegraphics[width=\linewidth]{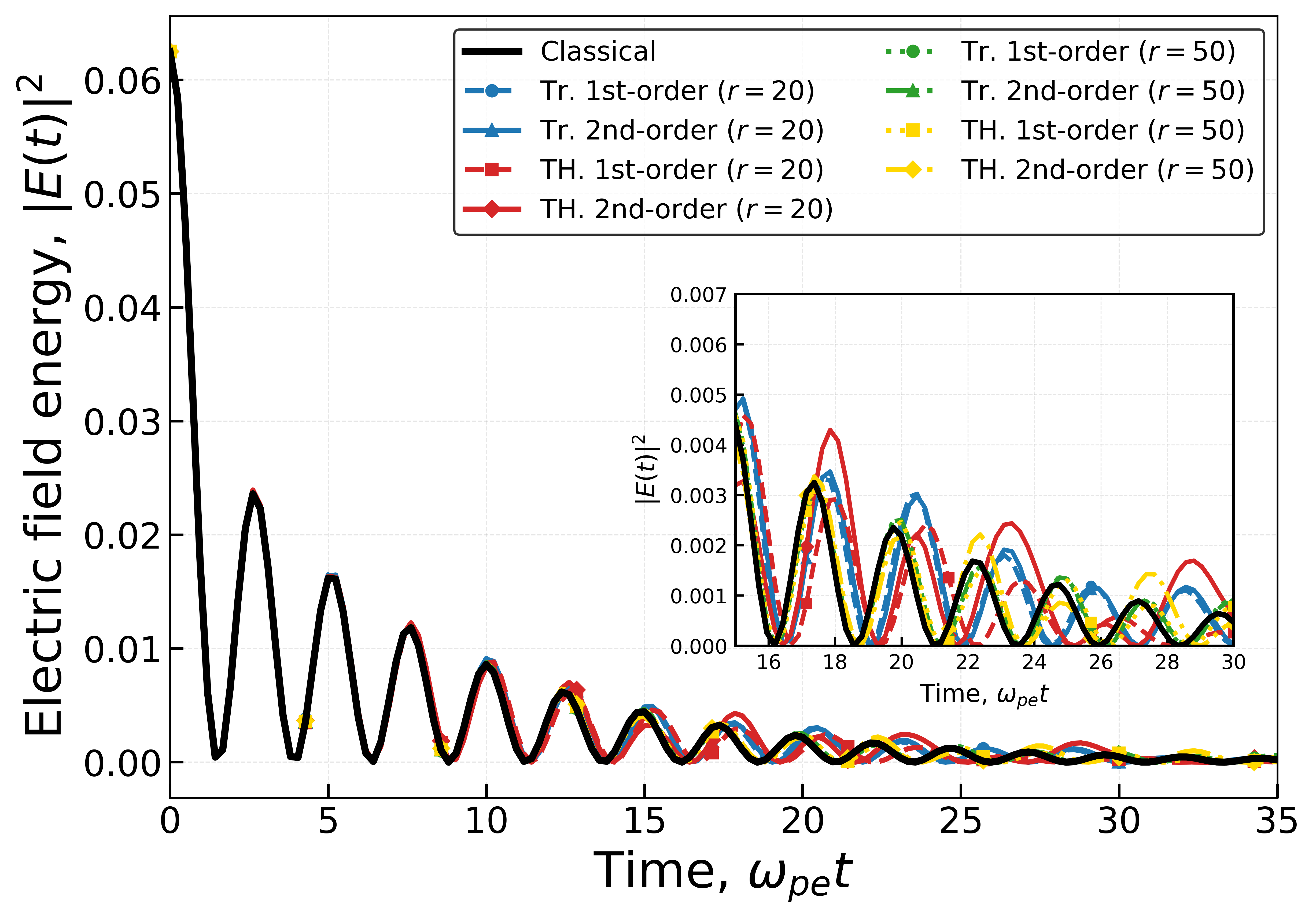}
    \caption{
    Temporal evolution of the electric-field energy, $|E(t)|^2$ for the non-magnetized ($\Omega_c=0$) linearized Vlasov-Poisson system with $k=0.45$, $N_v=31$,
    $v_{\max}=5.0\,v_T$ and $\Delta t=0.24$.
    The resulting $32$-dimensional Hamiltonian is encoded into a five-qubit register. The black solid curve represents the exact
    Hamiltonian evolution, while the colored curves show the first and second-order Trotter--Suzuki (Tr.) and THRIFT (TH.)
    approximations for $r=20$ and $r=50$ product-formula steps. The approximate solutions reproduce the oscillatory decay of the
    electric-field energy and converge toward the exact evolution as the number of product-formula steps and the approximation order increase
    are increased.In the inset, the long-range time version of the same plot is represents.
    }
    \label{fig:nonmag}
\end{figure}

Figure~\ref{fig:nonmag} shows the electric-field energy obtained from the exact Hamiltonian evolution together with the corresponding
Trotter--Suzuki and THRIFT approximations. For the parameters considered here, an exponential fit to the envelope of the electric-field energy, $\mathcal{E}{\mathrm{env}}(t)\propto e^{-2\gamma t}$, gives a damping rate of $\gamma=0.10907$. The corresponding dominant frequency yields a phase velocity of $v{\mathrm{ph}}=\omega/k=2.999,v_T$, where $v_T$ denotes the thermal velocity. The
characteristic oscillatory decay of the electric field amplitude is consistent with linear Landau damping in a collisionless plasma. The decay results from phase mixing between the perturbed particle distribution and the self-consistent electric field, causing the electric-field energy to decrease as the
perturbation develops increasingly fine structures in velocity space.

All four product-formula implementations reproduce the principal oscillation and the overall damping envelope of the exact solution.
The agreement demonstrates that the Hermitian Hamiltonian representation preserves the relevant linear kinetic dynamics under
the considered digital evolution. Small deviations from the exact solution become progressively more visible at later times ($\omega_{pe}t \gtrsim 15$). These
deviations originate from the finite product-formula step size and the non-commutativity of the Hamiltonian components.
\begin{table}[h]
\centering
\caption{
Absolute electric-field energy error,
$\Delta _E(t)=\left||E_{\mathrm{PF}}(t)|^2-|E_{\mathrm{exact}}(t)|^2\right|$,
for $r=20$ at selected evolution times.
}
\label{tab:error_r20}
\begin{tabular}{ccccc}
\hline
$\omega_{pe}t$ & Tr. 1st & Tr. 2nd & TH. 1st & TH. 2nd \\
\hline
1  & $6.208\times10^{-4}$ & $1.455\times10^{-4}$ & $1.561\times10^{-4}$ & $1.857\times10^{-5}$ \\
5  & $1.164\times10^{-3}$ & $7.803\times10^{-4}$ & $5.004\times10^{-4}$ & $6.021\times10^{-4}$ \\
10 & $2.795\times10^{-3}$ & $2.437\times10^{-3}$ & $9.329\times10^{-4}$ & $1.246\times10^{-3}$ \\
15 & $3.311\times10^{-3}$ & $2.221\times10^{-3}$ & $2.745\times10^{-3}$ & $9.833\times10^{-3}$ \\
25 & $1.109\times10^{-2}$ & $1.477\times10^{-2}$ & $3.759\times10^{-2}$ & $3.917\times10^{-2}$ \\
\hline
\end{tabular}
\end{table}
\begin{table}[h]
\centering
\caption{
Absolute electric-field energy error,
$\Delta_E(t)=\left||E_{\mathrm{PF}}(t)|^2-|E_{\mathrm{exact}}(t)|^2\right|$,
for $r=50$ at selected evolution times.
}
\label{tab:error_r50}
\begin{tabular}{ccccc}
\hline
$\omega_{pe}t$ & Tr. 1st & Tr. 2nd & TH. 1st & TH. 2nd \\
\hline
1  & $2.482\times10^{-4}$ & $5.902\times10^{-5}$ & $6.073\times10^{-5}$ & $7.035\times10^{-6}$ \\
5  & $3.821\times10^{-4}$ & $2.272\times10^{-4}$ & $1.841\times10^{-4}$ & $2.779\times10^{-4}$ \\
15 & $1.423\times10^{-3}$ & $1.551\times10^{-3}$ & $1.292\times10^{-3}$ & $2.923\times10^{-3}$ \\
25 & $4.279\times10^{-3}$ & $4.464\times10^{-3}$ & $3.969\times10^{-3}$ & $3.406\times10^{-3}$ \\
35 & $9.409\times10^{-3}$ & $9.285\times10^{-3}$ & $7.083\times10^{-3}$ & $7.925\times10^{-3}$ \\
\hline
\end{tabular}
\end{table}
For $r=20$, the agreement with the exact evolution becomes less accurate at later times, with the effect being more evident for the first-order approximations. Increasing the number of product-formula steps to $r=50$ improves the agreement for both Trotter--Suzuki and THRIFT. The second-order schemes also provide better agreement with the exact dynamics than the corresponding first-order schemes for most of the evolution. These results are shown in the inset of the Fig. \ref{fig:nonmag}. The reduction in error with increasing $r$ indicates improved convergence of the product-formula approximation toward the exact Hamiltonian evolution. The same behavior is reflected in the absolute electric-field energy errors listed in Tables~\ref{tab:error_r20} and \ref{tab:error_r50}.

In the present calculations,
$\alpha=1$ is used. Consequently, the observed differences between Trotter--Suzuki and THRIFT cannot be attributed solely to a
small-$\alpha$ suppression of the interaction term. Instead, they
arise from the different Hamiltonian decompositions and the
corresponding treatment of the dominant evolution. The results indicate that THRIFT can provide comparable or smaller observable errors for the considered plasma parameters, particularly at early times.

Overall, Fig.~\ref{fig:nonmag} and Tables~\ref{tab:error_r20} and \ref{tab:error_r50} demonstrate that both Trotter--Suzuki and THRIFT product formulas successfully reproduce the characteristic Landau-damped electric-field dynamics of the linearized Vlasov--Poisson system. The tabulated errors show that increasing the number of decomposition steps reduces the deviation from the exact Hamiltonian evolution, while higher-order product formulas generally provide improved accuracy. These results confirm the convergence of the digital simulations toward the exact dynamics and provide the basis for the detailed convergence and quantum-resource analysis presented in the following sections.
\subsection{Electric-Field Energy Evolution in a Magnetized Plasma}
We consider here, the linearized Vlasov--Poisson model in the presence of a uniform external magnetic field. The magnetic contribution
introduces velocity space rotation and couples the two transverse velocity components, thereby modifying the collisionless electric-field
response\cite{Bell2026}. To characterize this dynamics, we use the electric-field
energy
\begin{equation}
\mathcal{E}(t)=|E_x(t)|^2+|E_y(t)|^2 .
\label{eq:magnetic_energy}
\end{equation}
The exact dynamics are obtained by diagonalizing the Hermitian Hamiltonian as described in the previous section. The electric-field energy exhibits damped oscillations with a progressively decaying envelope, reflecting the transfer of coherent field energy into fine-scale particle dynamics through collisionless phase mixing. The decrease in electric-field energy does not imply dissipation, since the total evolution remains unitary and the Hamiltonian norm is conserved.

\begin{figure}[h]
    \centering
    \includegraphics[width=\linewidth]{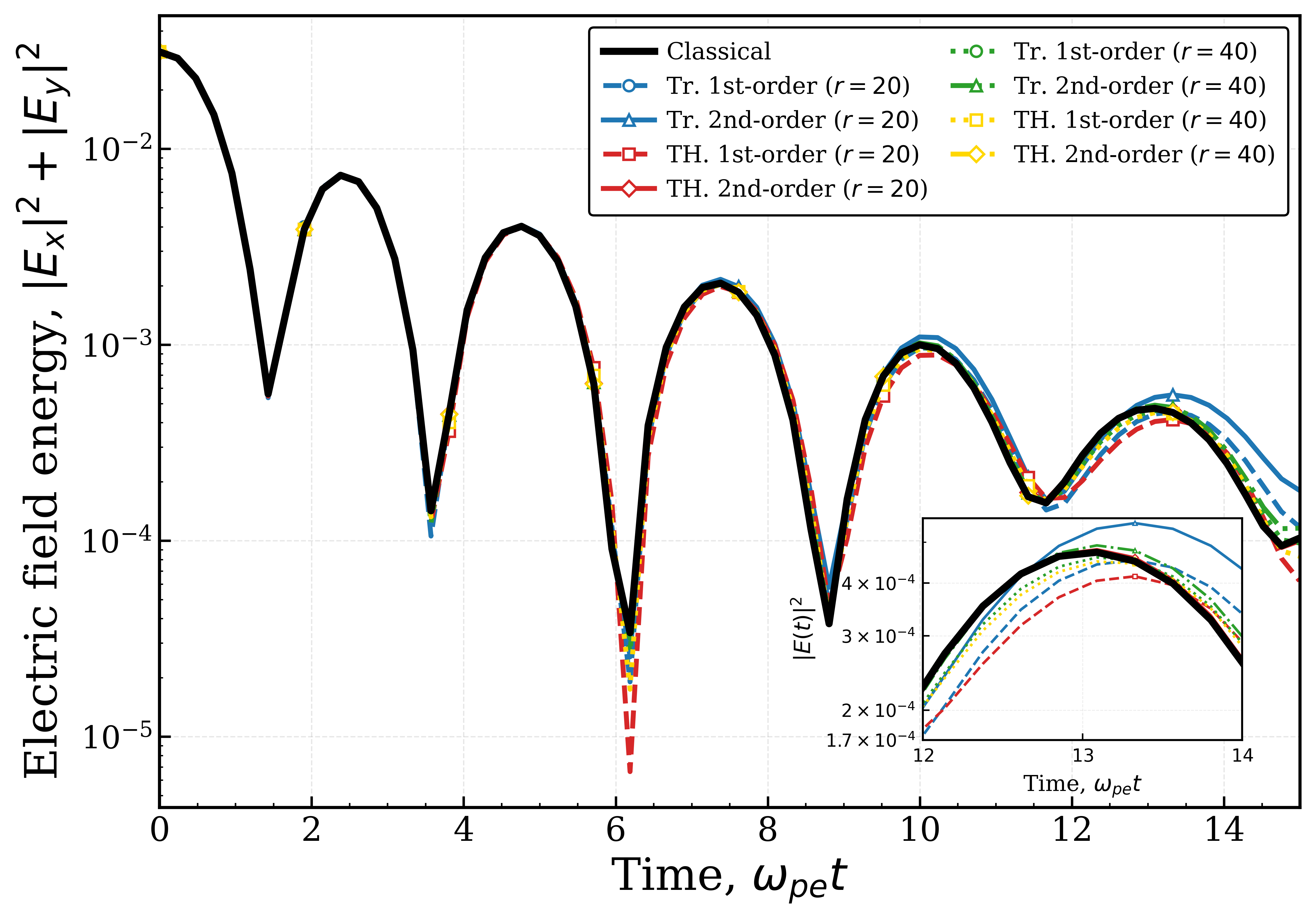}
    \caption{
Semi-logarithmic evolution of the normalized electric-field energy,
$\mathcal{E}(t)=|E_x(t)|^2+|E_y(t)|^2$, for the magnetized linearized Vlasov--Poisson system with $k_x=k_y=0.4$ and $B_0=0.3$.
A $31\times31$ velocity-space discretization is used, giving$31^2+2=963$ physical degrees of freedom and a $1024$-dimensional padded Hilbert space encoded by ten qubits. The black solid curve denotes the exact Hamiltonian evolution, while the colored curves show the first- and second-order Trotter--Suzuki (Tr.) and THRIFT (TH.) simulations for $r=20$ and $r=40$ decomposition steps.
}
\label{fig:mag_energy}
\end{figure}

Figure~\ref{fig:mag_energy} shows the resulting electric-field energy for $k_x=k_y=0.4$ and $B_0=0.3$. The product-formula approximations closely reproduce the exact electric-field dynamics. The deviations are more pronounced for the first-order formulas with $r=20$, particularly at later evolution
times ($\omega_{pe}t \approx 14$), whereas the second-order formulas provide improved agreement.
Increasing the number of product-formula steps from $r=20$ to $r=40$
further reduces the finite-step error for both Trotter--Suzuki and
THRIFT. These results indicate that first-order product formulas can
provide sufficient accuracy for short-time evolution when an appropriate number of decomposition steps is used. For longer
evolution times, a larger value of $r$ is required to maintain the same level of accuracy. The systematic improvement with increasing
$r$ and product-formula order confirms the expected convergence toward the exact Hamiltonian dynamics.

To quantify the damping of the electric-field energy, the maxima of $\mathcal{E}(t)$ are used to determine its decaying envelope. The
envelope is fitted to
\begin{equation*}
\mathcal{E}_{\rm env}(t)\propto e^{-2\gamma t},
\end{equation*}
where $\gamma$ is the damping rate of the electric-field amplitude.
The quality of the exponential fit is quantified by the coefficient of
determination, $R^2$.
\begin{table}[h]
\centering
\caption{
Damping and frequency characteristics of the electric-field dynamics
for different wave numbers and magnetic-field strengths. Here, $v_T=1$, $v_{\rm ph}$ is reported in units of $v_T$ and
$\mathcal{E}(0)$ denotes the initial electric-field energy at $t=0$.
}
\label{tab:magnetic_parameters}
\begin{tabular}{ccccccccc}
\hline
$k_x$ & $k_y$ & $B_0$ & $\gamma$ & $R^2$
& $f_{\rm peak}$ & $\omega$ & $v_{\rm ph}/v_T$
& $\mathcal{E}(0)$ \\
\hline
0.3 & 0.3 & 0.0 & 0.0860 & 0.9997
& 0.2099 & 1.3190 & 3.1089 & $5.56\times10^{-2}$ \\

0.4 & 0.4 & 0.0 & 0.2240 & 1.0000
& 0.2387 & 1.4996 & 2.6509 & $3.13\times10^{-2}$ \\

0.5 & 0.5 & 0.0 & 0.4033 & 0.9999
& 0.2688 & 1.6889 & 2.3885 & $2.00\times10^{-2}$ \\

0.3 & 0.3 & 0.3 & 0.0598 & 0.9923
& 0.1687 & 1.0599 & 2.4982 & $5.56\times10^{-2}$ \\

0.4 & 0.4 & 0.3 & 0.1291 & 0.9999
& 0.1794 & 1.1272 & 1.9927 & $3.13\times10^{-2}$ \\

0.5 & 0.5 & 0.3 & 0.2223 & 0.9959
& 0.1901 & 1.1944 & 1.6892 & $2.00\times10^{-2}$ \\
\hline
\end{tabular}
\end{table}
The extracted damping and frequency parameters are summarized in
Table~\ref{tab:magnetic_parameters}. In the absence of a magnetic
field, the damping rate increases with increasing $k$. The introduction
of a magnetic field reduces the damping rate over the parameter range
considered, indicating that the magnetic field modifies the collisionless
damping dynamics. The dominant frequency is obtained from the Fourier
spectrum of the electric-field components and is used to determine the
phase velocity through
\begin{equation}
\omega=2\pi f_{\rm peak},
\qquad
v_{\rm ph}=\frac{\omega}{k}.
\end{equation}
The results show that the magnetic field also shifts the dominant
frequency and consequently modifies the phase velocity.

For the representative case $k_x=k_y=0.4$, the damping rate decreases
from $\gamma=0.2240$ for $B_0=0$ to $\gamma=0.1291$ for $B_0=0.3$,
while the phase velocity decreases from $2.6509\,v_T$ to
$1.9927\,v_T$. The corresponding damping fits give $R^2>0.99$,
indicating that the electric-field-energy envelope is well described
by the assumed exponential decay. These results demonstrate that the
magnetic field significantly modifies both the damping and propagation
characteristics of the electric-field perturbation.

Overall, the results presented in Fig.~\ref{fig:mag_energy} and
Table~\ref{tab:magnetic_parameters} demonstrate that the proposed Hermitian Hamiltonian formulation captures the essential features of
magnetized collisionless plasma dynamics. The exact evolution exhibits the expected damped oscillatory behavior of the electric-field energy, while the extracted damping rates and dominant frequencies quantify the
effects of the wave number and magnetic-field strength. In particular, the results show that the magnetic field modifies both the damping rate
and the phase velocity of the electric-field perturbation.The Trotter--Suzuki and THRIFT approximations reproduce the main features
of the exact dynamics shown in Fig.~\ref{fig:mag_energy}. Their agreement
improves with increasing product-formula order and number of
decomposition steps, demonstrating systematic convergence toward the
exact evolution. These results establish the applicability of the
proposed quantum-simulation framework to magnetized plasma dynamics and
provide the basis for the comparison of the accuracy and quantum
resources of the Trotter--Suzuki and THRIFT approaches presented in the
following sections.
\subsection{Convergence of Trotter and THRIFT  Approximations}
\label{sec:convergence}
To quantitatively assess the accuracy of the product-formula approximations,
we consider the global state vector error measured by the Euclidean ($L_2$)
norm. The exact state at time $t$ is obtained by evolving the initial state
under the full plasma Hamiltonian $H$ given in Eq.~\eqref{HP},
\begin{equation}
|\psi_{\mathrm{exact}}(t)\rangle
=
e^{-iHt}|\psi(0)\rangle .
\label{eq:exact_state}
\end{equation}\\
The state obtained from a product-formula approximation with $r$
decomposition steps is denoted by
$|\psi_{\mathrm{approx}}(t;r)\rangle$. The corresponding global
state vector error is defined as
\begin{equation}
\delta_{\mathrm{state}}(r)
=
\left\|
|\psi_{\mathrm{exact}}(t)\rangle
-
|\psi_{\mathrm{approx}}(t;r)\rangle
\right\|_2 ,
\label{eq:state_error}
\end{equation}
where the Euclidean ($L_2$) norm of a complex vector
$|\phi\rangle=(\phi_1,\phi_2,\ldots,\phi_N)^T$ is given by
\begin{equation}
\left\| |\phi\rangle \right\|_2
=
\sqrt{
\sum_{j=1}^{N}|\phi_j|^2
}.
\label{eq:l2_norm}
\end{equation}
Consequently, $\delta_{\mathrm{state}}(r)$ measures the Euclidean distance
between the exact and approximate state vectors. A smaller value of
$\delta_{\mathrm{state}}(r)$ indicates closer agreement with the exact
Hamiltonian evolution. By evaluating this error for different values of
$r$, the convergence of the product-formula approximations can be
quantitatively assessed \cite{Suzuki1990,Childs2021}. The errors reported in this work are evaluated directly from state-vector simulations, thereby providing a quantitative measure of the convergence of the digital
Hamiltonian-evolution algorithms.

For a product formula of order $m$, the leading global discretization
error is expected to decrease with the number of decomposition steps $r$
according to
\begin{equation}
\epsilon(r)
\sim
C_m(t,\alpha)\,r^{-m},
\label{eq:error_scaling}
\end{equation}
where $C_m(t,\alpha)$ is a method and problem dependent prefactor.
For fixed evolution time and interaction strength, an $m$th-order product
formula therefore exhibits the asymptotic convergence rate
$\mathcal{O}(r^{-m})$. In particular, first, second and fourth-order
product formulas are expected to exhibit $r^{-1}$, $r^{-2}$, and $r^{-4}$
convergence, respectively \cite{Childs2021}.Taking the logarithm of Eq.~\eqref{eq:error_scaling} gives
\begin{equation}
\log \epsilon(r)
=
-m\log r
+
\log C_m(t,\alpha),
\label{eq:log_error_scaling}
\end{equation}
which shows that the slope of a log--log plot of the error versus the
number of decomposition steps approaches $-m$ in the asymptotic regime.
Thus, the numerical slope provides a direct measure of the observed
convergence order.

For THRIFT, the same order of convergence is expected because the
underlying product formula is still of order $m$. The corresponding error
can therefore be characterized by the same scaling relation in
Eq.~\eqref{eq:error_scaling}, with a THRIFT-specific prefactor
$C_m^{\mathrm{THRIFT}}(t,\alpha)$. The interaction strength $\alpha$
primarily affects this prefactor rather than the asymptotic power of $r$.
In the interaction-picture formulation used by THRIFT, the dominant
part of the Hamiltonian is treated separately, leaving a reduced
residual interaction. This can decrease the magnitude of the error
prefactor while preserving the expected $r^{-m}$ convergence order
\cite{Bosse2025}. So, $\alpha$ affects the magnitude of
the approximation error through the prefactor but does not alter the
formal convergence order $m$.

\begin{figure}[h]
\centering
\includegraphics[width=\linewidth]{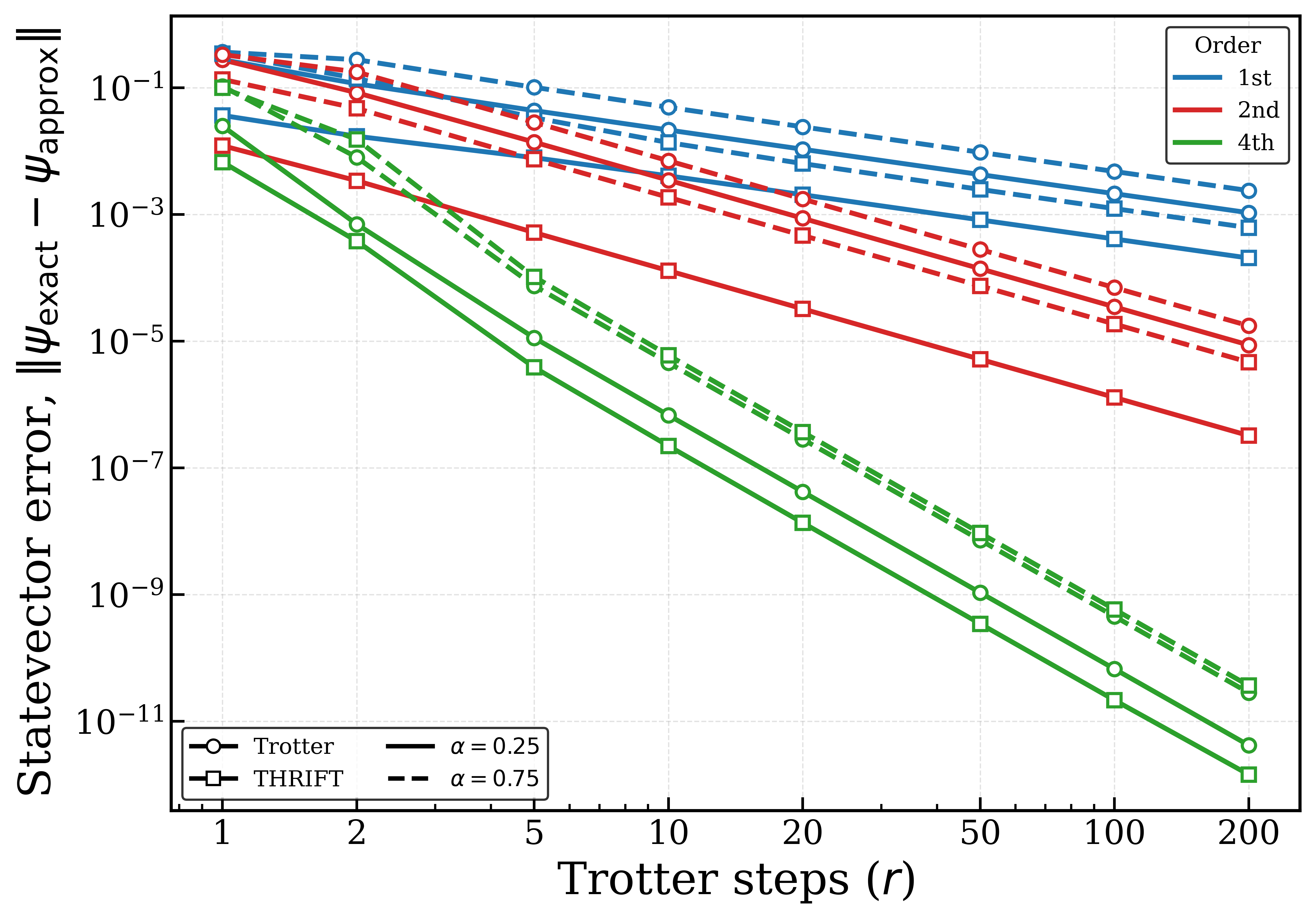}
\caption{State-vector error, $\|\psi_{\mathrm{exact}}(5)-\psi_{\mathrm{approx}}(5)\|_{2}$,as a function of the number of product-formula steps $r$ for the first, second and fourth-order Trotter and THRIFT algorithms, evaluated at the fixed evolution time $t=5$. Solid and dashed curves correspond to interaction strengths $\alpha=0.25$ and $\alpha=0.75$,respectively.}
\label{fig:error_scaling}
\end{figure}

Figure~\ref{fig:error_scaling} presents the state vector error as a function of the number of product-formula steps $r$ for the first, second and fourth-order Trotter and THRIFT schemes. The log-log representation allows the asymptotic convergence rates to be assessed directly from the slopes. For all schemes, the error decreases systematically with increasing $r$, while Higher-order formulas show faster convergence as the number of decomposition steps increase. In particular, the fourth-order schemes achieve a markedly faster reduction in error than the corresponding first and second-order approximations, consistent with the theoretical scaling in Eqs.~\eqref{eq:error_scaling} and~\eqref{eq:thrift_scaling}.
To quantitatively assess the observed convergence rates, a linear least-squares fit was performed in the $\log\epsilon$ vs $\log r$ plane using the last five data points, which represent the large $r$ asymptotic regime. The fitted slopes provide a numerical estimate of the convergence order. For $N_v=31$ at the fixed evolution time $t=5$, the resulting slopes are summarized in Table~\ref{tab:convergence_slopes}.
\begin{table}[h]
\centering
\caption{
Log-log convergence of the Trotter and THRIFT product formulas for
$N_v=31$ and $t=5$. Slopes are fitted using the last five data points.
}
\label{tab:convergence_slopes}
\begin{tabular}{c cc cc cc}
\hline
& \multicolumn{2}{c}{1st order}
& \multicolumn{2}{c}{2nd order}
& \multicolumn{2}{c}{4th order} \\
\cline{2-7}
$\alpha$
& Trotter & THRIFT
& Trotter & THRIFT
& Trotter & THRIFT \\
\hline
0.25
& $-1.0033$ & $-0.9973$
& $-1.9996$ & $-2.0001$
& $-4.0036$ & $-3.9907$ \\

0.50
& $-1.0057$ & $-1.0071$
& $-2.0004$ & $-1.9998$
& $-4.0040$ & $-4.0062$ \\

0.75
& $-1.0098$ & $-1.0297$
& $-2.0011$ & $-1.9995$
& $-4.0014$ & $-4.0059$ \\

1.00
& $-1.0199$ & $-1.0379$
& $-2.0009$ & $-1.9992$
& $-3.9984$ & $-3.9904$ \\
\hline
\end{tabular}
\end{table}
The small deviations from the ideal slopes indicate that the numerical results have reached the expected asymptotic regime. Moreover, the weak variation of the fitted slopes with $\alpha$ demonstrates that the interaction-strength parameter does not alter the formal convergence order. Instead, its effect is reflected primarily in the method-dependent error prefactor $C_m(t,\alpha)$ in Eq.~(\ref{eq:error_scaling}). Thus, the convergence order describes how the error decreases as the
number of product-formula steps increases, while the absolute error at a fixed $r$ provides a direct measure of the effect of $\alpha$ and allows comparison of the accuracy of Trotter and THRIFT.
\subsection{Two-Qubit Gate Depth}
In addition to the approximation error, we evaluate the quantum circuit resources required by the Trotter and THRIFT product-formula methods. Circuit depth is an important resource measure for Hamiltonian simulation because it determines the number of sequential layers of operations required to execute a quantum circuit, while operations acting on disjoint qubits can be performed in parallel \cite{Clinton2021,JavadiAbhari2024,Qiskit2024}. In particular, we focus on the two-qubit gate
depth, which characterizes the sequential cost of entangling
operations.\\
For a circuit, we define the two-qubit gate depth as
\begin{equation}
D_{2q}
=
\operatorname{depth}\!\left(C\big|_{\mathrm{2q}}\right),
\label{eq:qiskit_depth}
\end{equation}
where $C\big|_{\mathrm{2q}}$ denotes the circuit restricted to two-qubit operations. Thus, $D_{2q}$ counts the number of sequential layers containing two-qubit gates, while allowing two-qubit gates acting on different qubit pairs to be executed in parallel. This quantity is different from the total number of two-qubit gates, which counts every entangling operation independently of whether it can be executed in parallel.

In the present calculations, $D_{2q}$ is evaluated directly from the
Qiskit circuit representation by applying the circuit-depth function
with a filter that retains only two-qubit operations. The corresponding
two-qubit gate count is obtained independently from the circuit
operation counts \cite{Qiskit2024}.

\begin{figure}[h]
    \centering
    \includegraphics[width=\linewidth]{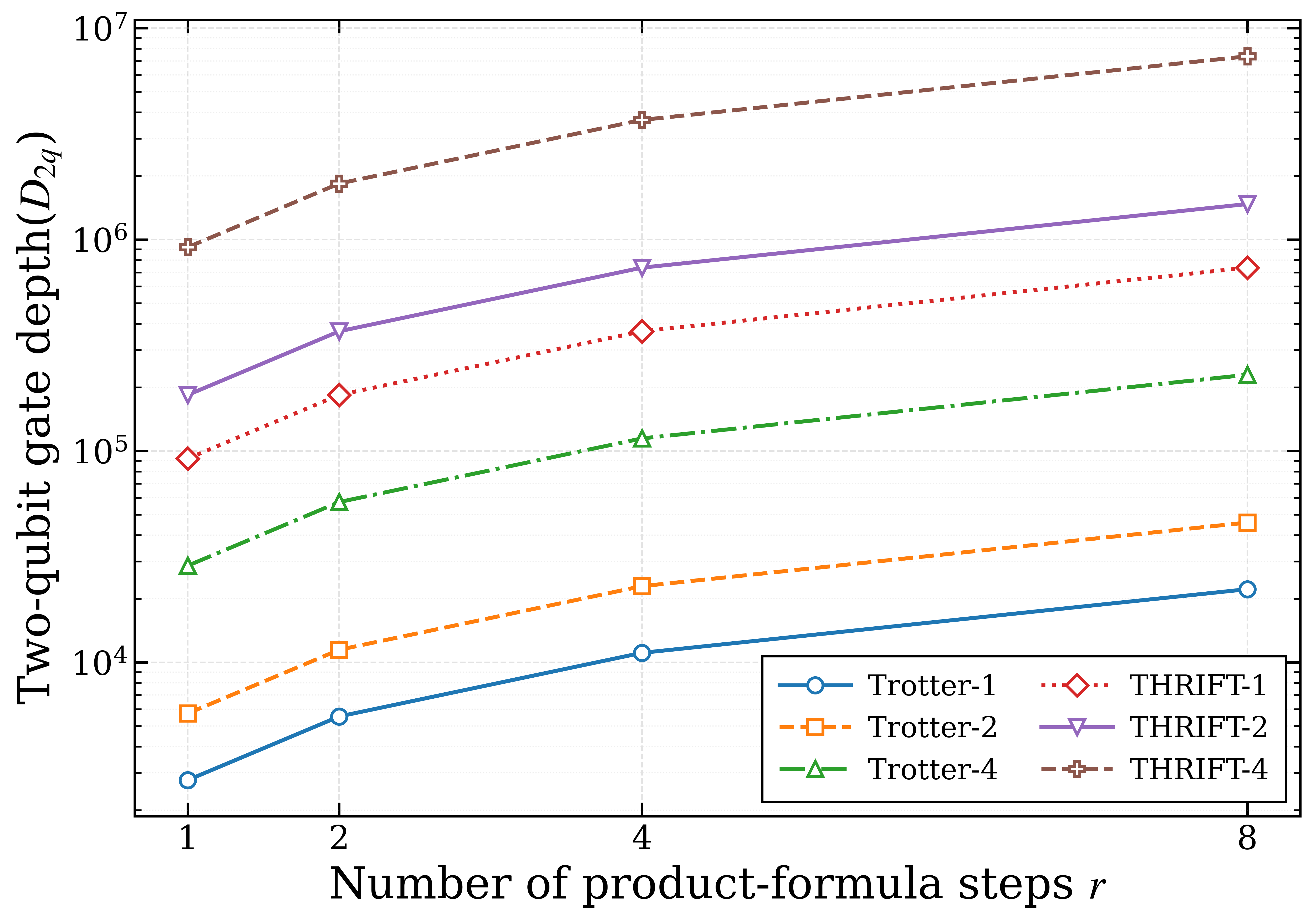}
    \caption{Two-qubit gate depth $D_{2q}$ versus the number of product-formula steps $r$ for first, second and fourth-order Trotter and THRIFT circuits. The vertical axis $D_{2q}$ is plotted on a logarithmic scale. The calculations use the discretized Vlasov--Poisson Hamiltonian with $N_v=31$ (five qubits) and a total evolution time $t=1$. The two-qubit gate depth is calculated by retaining only two-qubit operations when determining the circuit depth.}
    \label{fig:depth_compare}
\end{figure}

Figure~\ref{fig:depth_compare} shows the two-qubit gate depth as a
function of the number of product-formula steps $r$ for the first
second and fourth-order Trotter and THRIFT schemes.
For a fixed product-formula order, the two-qubit gate depth increases
approximately linearly with the number of product-formula steps,
\[
D_{2q}^{(m)}(r) \propto r.
\]
This behavior results from the repeated application of the elementary product-formula sequence. At a fixed value of $r$, higher order formulas generally exhibit a larger two-qubit gate depth because their
construction involves a larger number of elementary evolution
operations.\\
The THRIFT circuits exhibit a larger two-qubit gate depth per product-formula step than the corresponding Trotter circuits in the
present implementation. This additional cost is associated with the interaction-picture operations used in the THRIFT construction. Therefore, the two-qubit depth at a fixed $r$ does not by itself determine the overall efficiency of a method.

\begin{figure}[h]
\centering
\includegraphics[width=\linewidth]{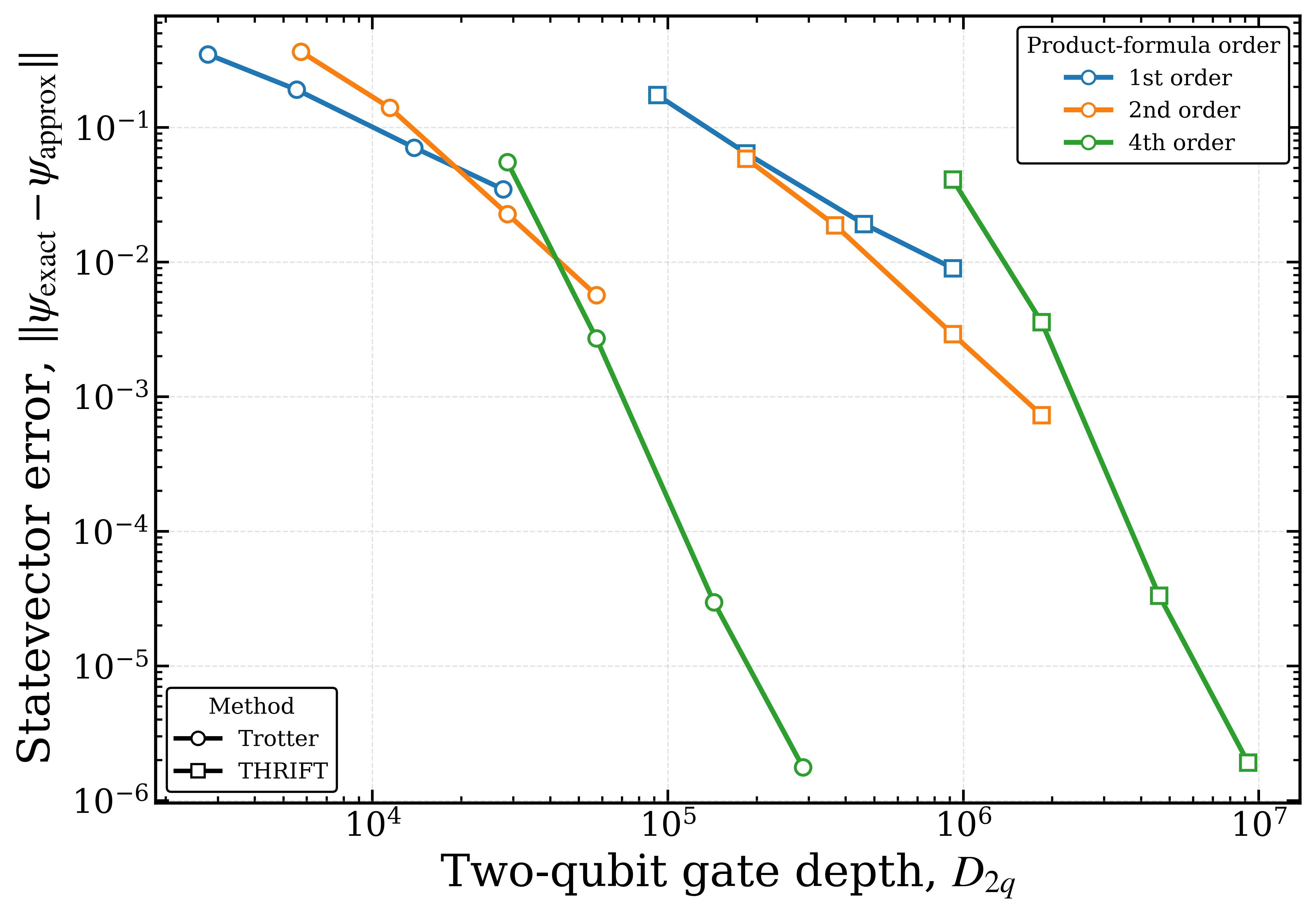}
\caption{ State-vector error as a function of the two-qubit gate depth $D_{2q}$ for first, second and fourth-order Trotter and THRIFT product formulas. The logarithmic axes show the reduction in approximation error with increasing two-qubit circuit resources. Higher order product formulas achieve substantially smaller errors at comparable two-qubit gate depths. } \label{fig:error_depth}
\end{figure}

A more meaningful comparison is obtained by relating the circuit depth directly to the simulation accuracy. Figure~\ref{fig:error_depth} shows the state-vector error ($\delta_{\mathrm{state}}$) as a function of the two-qubit gate depth.
This behavior is consistent with their higher asymptotic convergence order. The THRIFT results follow the same general trend, while the required two-qubit depth is larger than that of the corresponding Trotter circuits in the present implementation.

The error depth representation provides a more useful measure of overall simulation efficiency than the depth at a fixed number of product-formula steps. For a prescribed target accuracy, the relevant quantity is the minimum two-qubit gate depth required to achieve that accuracy. So, a method with a larger circuit cost per product-formula step can still be competitive if its faster error convergence reduces the number of steps required to reach the target accuracy. The results therefore demonstrate a trade-off between the cost of an individual product-formula step and the accuracy gained from increasing the product-formula order.

%\subsection{Entanglement bound on the geometrically local circuit cost}
The preceding analysis quantifies the two-qubit resources of the particular Trotter and THRIFT circuit constructions used in our simulations. These gate counts should not, however, be interpreted as the minimum circuit complexity of the simulated evolution. To place the observed gate counts in a broader physical context, we next consider an entanglement-based lower bound on the minimum number of local two-qubit gates required to generate the evolved state. Exact calculation of the gate depth is tremendously difficult \cite{Peterson2020TwoQubitDepth}.
The minimum number of two-qubit gates required to perform a unitary operation is called the quantum circuit complexity. This is also referred to as the gate complexity or the geometrically local circuit cost. But the circuit depth is the number of sequential layers applied in parallel. 
For a lattice of $L$ qubits, let $U(t)=e^{-iHt}$ be the unitary evolution generated by a local Hamiltonian. For a physical quantum circuit, there should be many quantum circuits that can create the desired unitary evolution, so optimizing this and finding the minimum number of two-qubit gates will be an exponentially hard task. Instead of evaluating the exact circuit cost, Eisert showed that it is possible to derive a rigorous lower bound using the entanglement generated during the unitary evolution~\cite{Eisert2021}.
Consider an initial product state $|\phi\rangle$. After the evolution,
\begin{equation}
|\psi(t)\rangle=U(t)|\phi\rangle,
\end{equation}
The bipartite entanglement entropy across the partition
$A=\{1,\cdots,\ell\}$ and$B=\{\ell+1,\cdots,L\}$ is
\begin{equation}
S_\ell
=
-
\mathrm{Tr}
\left(
\rho_A
\ln\rho_A
\right),
\end{equation}where
\begin{equation}
\rho_A
=
\mathrm{Tr}_B
\left(
|\psi(t)\rangle
\langle\psi(t)|
\right).
\end{equation}
Geometrical Entanglement Entropy(GEC) is defined as summing over the entanglemnt entropy over all the possible bipartition
\begin{equation}
E_g(U)
=
\frac{1}{\log2}
\sum_{\ell=1}^{L-1}
S_\ell.
\end{equation}
Using the upper bound on the entangling rate of local Hamiltonians derived by Bravyi~\cite{Bravyi2007}, together with the geometric formulation of quantum computation introduced by Nielsen \textit{et al.}~\cite{Nielsen2006}, Eisert proved that the geometrically local circuit cost satisfies the inequality
\begin{equation}
C_g(U)
\ge
\frac{1}{\mathcal{L}}
E_g(U),
\label{eq:complexity}
\end{equation}
where $\mathcal{L
}$ is a positive constant of order unity~\cite{Eisert2021}. Equation~(\ref{eq:complexity}) shows that the entanglement generated during the evolution provides a rigorous lower bound on the total number of local two-qubit gates required to implement the unitary. How can we say this is the lower bound? A local gate only generates a finite amount of entanglement across a particular bipartition, so many-body states require many local entangling gates.
\begin{figure}[h]
    \centering
    \includegraphics[width=\linewidth]{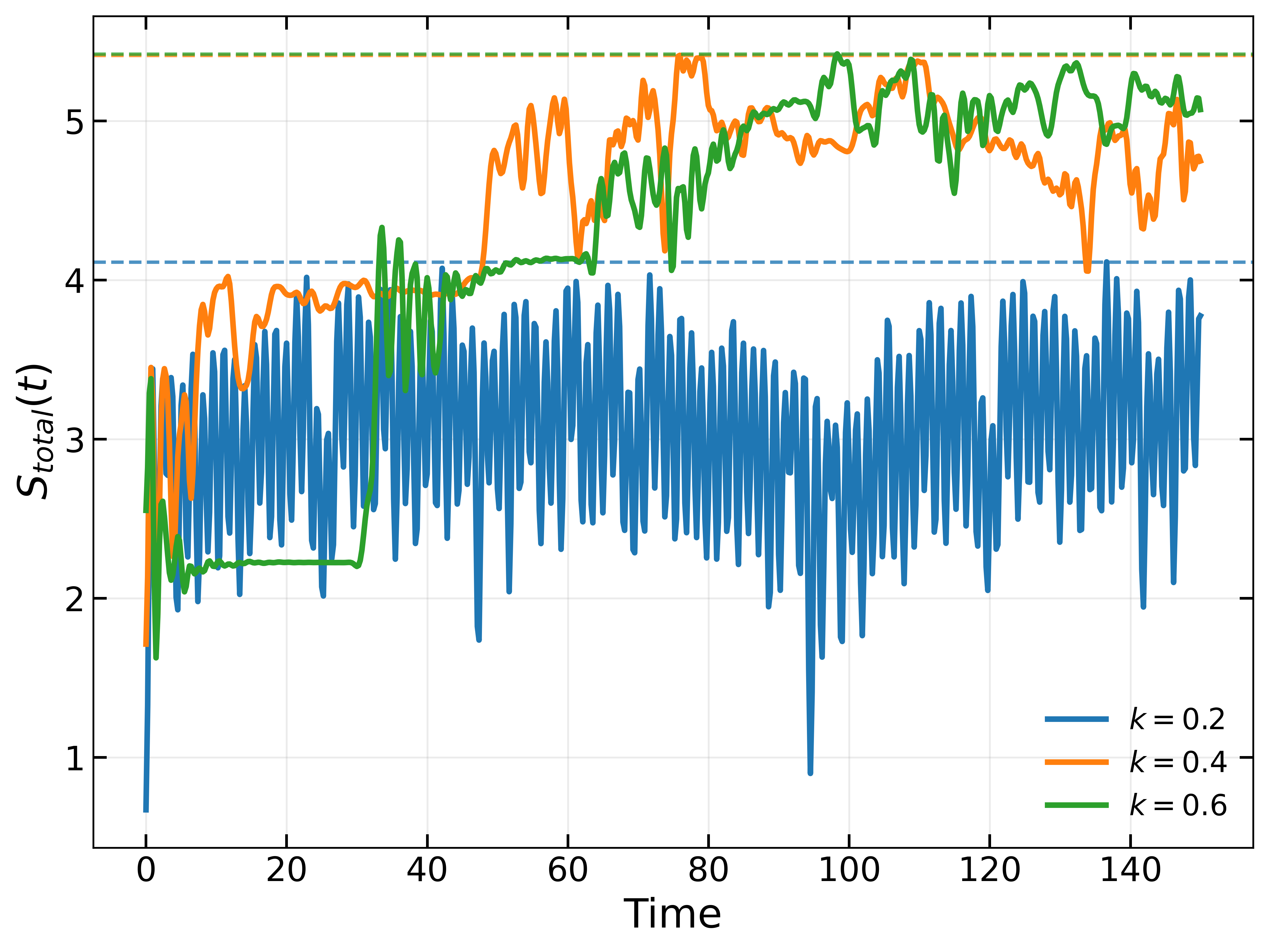}
    \caption{Time evolution of the total entanglement entropy, $S_{\mathrm{total}}(t)$, obtained by summing the von Neumann entropy over all bipartitions of the quantum register. The dashed horizontal line indicates the theoretical maximum entanglement entropy.}
    \label{fig:entanglement}
\end{figure}
Figure~\ref{fig:entanglement} shows the evolution of the total entanglement entropy during the Hamiltonian simulation. Starting from a weakly entangled initial state, we find that the entropy increases rapidly at early times, indicating the efficient generation of quantum correlations as the system evolves under the Vlasov--Poisson Hamiltonian. This initial growth is the result of the redistribution of quantum information among the qubits via the coherent dynamics of the simulation.\\
After the transient growth the entanglement entropy reaches a quasi-steady regime close to the theoretical maximum with persistent oscillations throughout the rest of the evolution . These oscillations are due to the coherent unitary evolution of the closed quantum system and show that quantum correlations still change without a substantial loss of coherence. Further evidence that the evolution of the simulation is unitary over the whole time interval is the absence of any systematic decay in the entropy.\\
The rapid generation and later saturation of entanglement indicate that the Hamiltonian is efficiently exploring the available Hilbert space while maintaining coherent quantum dynamics. The closeness of the entropy to its theoretical upper bound suggests the formation of highly entangled many-qubit states that are difficult to represent efficiently with classical methods. These results demonstrate that the proposed quantum simulation can faithfully reproduce the dynamical evolution of the Vlasov-Poisson system and the corresponding creation of multipartite quantum correlations.

 The entanglement-based lower bound and the actual gate counts obtained
from the Trotter and THRIFT circuits characterize different aspects of
the simulation cost. The lower bound captures the fundamental
entangling resources required by the dynamics, whereas the actual gate
count corresponds to a particular circuit construction and therefore
includes additional overhead from the product-formula decomposition and
the implementation of the individual Hamiltonian terms. Thus, the
substantial difference between the theoretical lower bound and the
realized gate count is expected and does not indicate a discrepancy in
the analysis. Rather, the lower bound provides a fundamental benchmark
against which the efficiency of the explicit circuit constructions can
be assessed. In addition, this bound concerns the minimum number of
local two-qubit gates and should not be directly identified with the
two-qubit gate depth $D_{2q}$, which also depends on the extent to which
the gates can be executed in parallel. Together, the entanglement
analysis and the circuit-depth results therefore distinguish the
intrinsic entangling cost of the dynamics from the practical circuit
overhead of the Trotter and THRIFT implementations.

\subsection{Error Mitigation via Richardson Extrapolation}
\label{sec:richardson}
The product-formula approximation introduces a systematic finite-step error that decreases as the number of product-formula
steps $r$ is increased. Instead of reducing this error solely by increasing the circuit depth, Richardson extrapolation (RE) can be used to cancel the leading contribution through a classical linear combination of simulations performed at different step
sizes \cite{Richardson1911,Temme2017,Krebsbach2022,Cai2023}.\\
For a product formula of order $p$, the state produced with $r$ steps can be expressed asymptotically as
\begin{equation}
|\psi_r\rangle
=
|\psi_{\mathrm{exact}}\rangle
+
C_p r^{-p}
+
\mathcal{O}\!\left(r^{-p-1}\right),
\label{eq:state_error_expansion}
\end{equation}
where $C_p$ denotes a method and problem-dependent error coefficient. Evaluating the same product formula with $2r$ steps
gives
\begin{equation}
|\psi_{2r}\rangle
=
|\psi_{\mathrm{exact}}\rangle
+
C_p(2r)^{-p}
+
\mathcal{O}\!\left(r^{-p-1}\right).
\end{equation}
The leading $\mathcal{O}(r^{-p})$ contribution can therefore be cancelled by the two-point Richardson extrapolation
\begin{equation}
|\psi_{\mathrm{RE}}\rangle
=
\frac{2^p|\psi_{2r}\rangle-|\psi_r\rangle}
{2^p-1}.
\label{eq:Richardson}
\end{equation}
Substitution of the above asymptotic expansions shows that the leading finite-step error cancels, leaving contributions from
higher-order terms in the error expansion. 
\begin{figure}[h!]
    \centering
    \includegraphics[width=\linewidth]{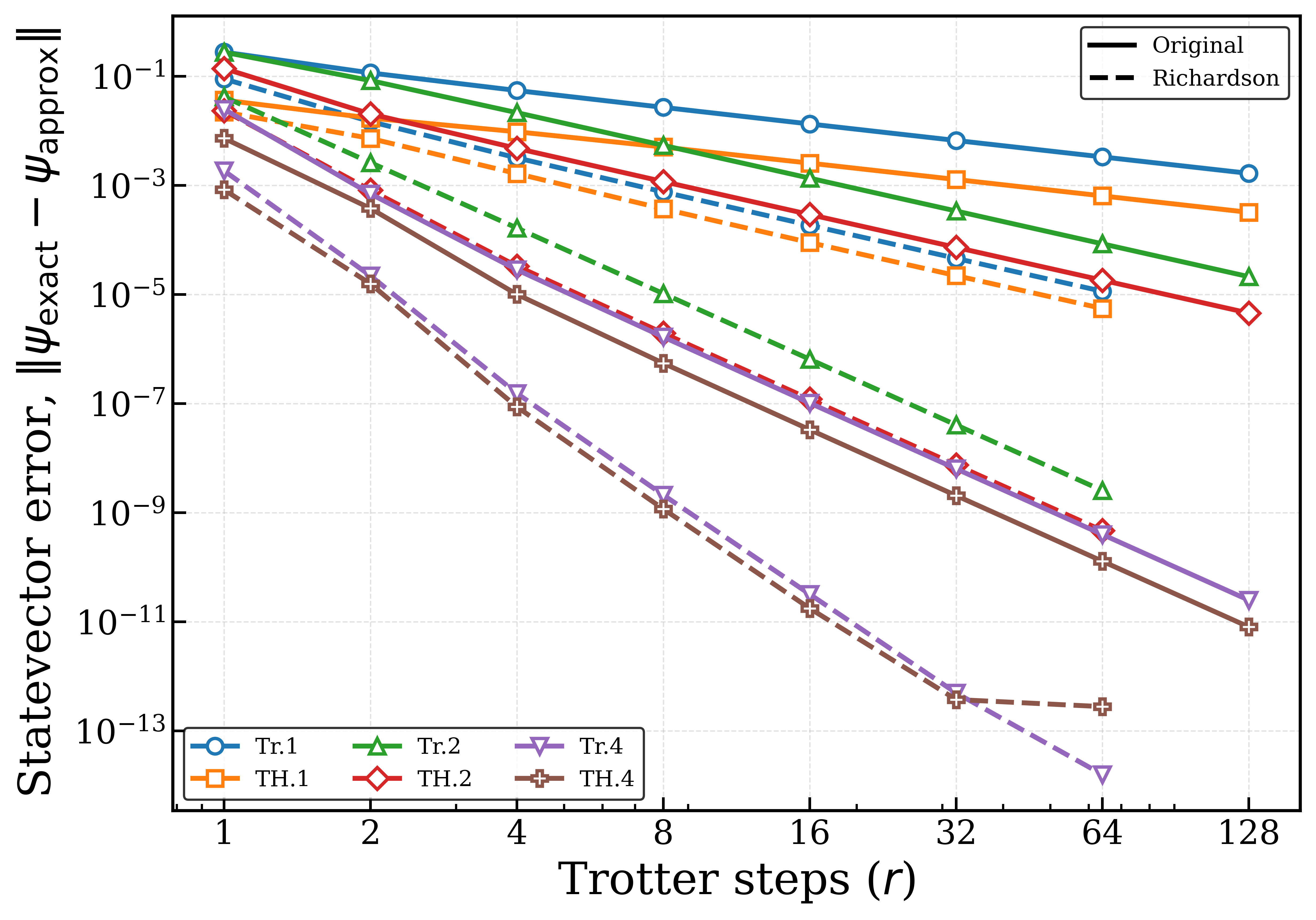}
    \caption{
State-vector error
$\|\psi_{\mathrm{exact}}-\psi_{\mathrm{approx}}\|_2$
as a function of the number of product-formula steps $r$ for the
first, second and fourth-order Trotter (Tr.) and THRIFT (TH.)
algorithms. Solid curves show the original product-formula results,whereas dashed curves show the corresponding results after
two-point Richardson extrapolation. The extrapolated states are constructed from simulations with $r$ and $2r$ steps using Eq.~\eqref{eq:Richardson}. The vertical axis is shown on a logarithmic scale.}
    \label{fig:richardson}
\end{figure}
Richardson extrapolation is therefore an error-reduction technique
that improves the accuracy by combining results obtained with
different numbers of simulation steps, without modifying the
individual quantum circuits \cite{Temme2017,Krebsbach2022}.
Figure~\ref{fig:richardson} compares the state-vector errors of the
original and Richardson-extrapolated Trotter and THRIFT
approximations. For all three approximation orders, Richardson extrapolation
systematically reduces the finite-step error over the range of
$r$ considered. The improvement results from the cancelation of
the leading error contribution in Eq.~\eqref{eq:Richardson}. The
effect is particularly pronounced for the first-order formulas,
where the original error decreases relatively slowly with $r$.
The second and fourth-order schemes also benefit from the
extrapolation, although their original discretization errors are
already smaller at comparable values of $r$.

For the fourth-order Trotter and THRIFT schemes, the error reaches
very small values at sufficiently large $r$. At this point, further
improvement is limited by numerical round-off and finite-precision
effects, leading to the observed saturation of the error at
approximately $10^{-13}$--$10^{-14}$. Thus, the departure from the
ideal convergence trend at the smallest errors should not be
interpreted as a failure of Richardson extrapolation, but rather as
the onset of numerical precision limits.The method provides a
trade-off between circuit depth and the number of circuit
evaluations: a more accurate result can be obtained without
increasing the depth of the individual circuits beyond that of the
largest $r$ simulation. This trade-off is central to the use of
Richardson extrapolation as a practical error-mitigation technique
for quantum simulations \cite{Temme2017,Krebsbach2022,Cai2023}.
\section{Conclusion}
We have investigated a quantum simulation framework for the linearised Vlasov–Poisson system, based on its Hermitian Hamiltonian formulation, to study dynamics in magnetized and unmagnetized plasma. The resulting dynamics were simulated using first, second and fourth-order Trotter--Suzuki and THRIFT product-formula methods. The numerical results demonstrate systematic convergence toward the exact Hamiltonian evolution with increasing product-formula order and number of decomposition steps. The electric-field dynamics are accurately reproduced for both non-magnetized and magnetized configurations, demonstrating the applicability of the Hamiltonian
formulation to linearized kinetic plasma dynamics.

The state vector error analysis confirms the expected asymptotic convergence orders of the product-formula approximations. THRIFT
provides a reduction in the finite-step error by treating the dominant streaming contribution in the interaction picture, with the benefit being particularly evident for weaker interaction strengths. This improvement is accompanied by an increased two-qubit gate depth per product-formula step. The results therefore highlight the trade-off between approximation accuracy and quantum circuit resources, indicating that circuit depth per step alone is not sufficient to assess the overall efficiency of a Hamiltonian simulation method.
Richardson extrapolation further improves the accuracy by canceling the leading finite-step error through classical post-processing. Although the extrapolation does not increase the depth of an individual quantum circuit, it requires simulations at different product-formula step sizes and therefore introduces additional execution cost. Consequently, it provides a practical approach for reducing discretization errors when additional circuit evaluations are preferable to increasing the depth of individual circuits.

This results establish product-formula Hamiltonian simulation as a systematic approach for simulating linearized kinetic plasma dynamics. The comparison of Trotter and THRIFT methods provides a quantitative assessment of the relationship between convergence, interaction strength, and quantum circuit resources. The framework developed in this work can be extended to larger velocity-space discretizations and more general kinetic plasma models. 
It will be interesting to investigate whether the observed THRIFT advantage persists for more complex plasma configurations and to extend the framework to time-dependent magnetic fields and collisional kinetic models. Future studies will also consider realistic hardware noise, finite sampling and gate errors to assess the performance of these methods on quantum hardware.

\begin{acknowledgments}
We acknowledge funding support from the National Quantum Mission, an initiative of the Department of Science and Technology, Govt. of India. We also acknowledge the support provided by the Foundation for QC Innovation (FQCI), DST-NQM T-Hub at IISc Bengaluru, in facilitating this project.
\end{acknowledgments}

\appendix
\section{Product-Formula Error Analysis}
\label{app:trotter}
Digital quantum simulation approximates the exact time-evolution operator
\begin{equation}
U(t)=e^{-iHt},
\end{equation}
where the Hamiltonian is decomposed into two generally non-commuting Hermitian
operators,
\begin{equation}
H=A+B.
\end{equation}
Since $[A,B]\neq0$, the exponential cannot be factorized exactly. Instead,
product-formula approximations are employed based on the Lie--Trotter and
Suzuki decompositions \cite{Trotter1959,Suzuki1993,Childs2021}.
\subsection{First-Order Lie--Trotter Formula}
The first-order product formula is
\begin{equation}
U_{1}(t)
=
\left(
e^{-iA\Delta t}
e^{-iB\Delta t}
\right)^r,
\qquad
\Delta t=\frac{t}{r},
\end{equation}
where $r$ denotes the number of Trotter steps.
Using the Baker--Campbell--Hausdorff (BCH) expansion,
\begin{equation}
e^{X}e^{Y}
=
\exp
\left(
X+Y
+\frac12[X,Y]
+\mathcal{O}(\Delta t^{3})
\right),
\end{equation}
with
\[
X=-iA\Delta t,
\qquad
Y=-iB\Delta t,
\]gives
\begin{equation}
\begin{aligned}
e^{-iA\Delta t}e^{-iB\Delta t}
=\exp[
-i(A+B)\Delta t
-\frac{1}{2}[A,B]\Delta t^{2}
+\mathcal{O}(\Delta t^{3})
].
\end{aligned}
\label{eq:BCH}
\end{equation}
Applying this approximation over $r$ time steps yields
\begin{equation}
U_{1}(t)
=
e^{-iHt}
+
\mathcal{O}
\left(
\frac{t^{2}}{r}
\right),
\end{equation}
so that the global error scales as
\begin{equation}
\epsilon_{1}
=
\mathcal{O}
\left(
\frac{t^{2}}{r}
\right).
\end{equation}
\subsection{Second-Order Strang Splitting}
The symmetric second-order decomposition is
\begin{equation}
U_{2}(t)
=
\left(
e^{-iA\Delta t/2}
e^{-iB\Delta t}
e^{-iA\Delta t/2}
\right)^r .
\end{equation}
Because of its symmetric structure, the leading second-order commutator errors
cancel identically in the BCH expansion. Consequently,
\begin{equation}
U_{2}(t)
=
e^{-iHt}
+
\mathcal{O}
\left(
\frac{t^{3}}{r^{2}}
\right),
\end{equation}
and the global error becomes
\begin{equation}
\epsilon_{2}
=
\mathcal{O}
\left(
\frac{t^{3}}{r^{2}}
\right).
\end{equation}
\subsection{Fourth-Order Suzuki Formula}
Higher-order accuracy is obtained by recursively composing the second-order
operator according to Suzuki's construction,
\begin{equation}
S_{4}(\Delta t)
=
S_{2}(p\Delta t)
S_{2}(p\Delta t)
S_{2}((1-4p)\Delta t)
S_{2}(p\Delta t)
S_{2}(p\Delta t),
\end{equation}
where
\begin{equation}
p=
\frac{1}{4-4^{1/3}}.
\end{equation}
The coefficient $p$ is chosen such that the leading third-order error terms in
the BCH expansion cancel exactly, resulting in
\begin{equation}
S_{4}(\Delta t)
=
e^{-iH\Delta t}
+
\mathcal{O}(\Delta t^{5}).
\end{equation}
The complete fourth-order approximation is
\begin{equation}
U_{4}(t)
=
\left(
S_{4}(\Delta t)
\right)^r,
\qquad
\Delta t=\frac{t}{r},
\end{equation}
which yields the global error
\begin{equation}
\epsilon_{4}
=
\mathcal{O}
\left(
\frac{t^{5}}{r^{4}}
\right).
\end{equation}
Thus, increasing the order of the product formula systematically suppresses the
Trotter error without modifying the underlying Hamiltonian. The first-,
second-, and fourth-order formulas therefore exhibit global convergence rates
of $\mathcal{O}(t^{2}/r)$,
$\mathcal{O}(t^{3}/r^{2})$, and
$\mathcal{O}(t^{5}/r^{4})$, respectively.
\section{Interaction-Picture THRIFT Formulation}
\label{app:thrift}
The Time-Resolved Interaction Framework (THRIFT) is based on separating the
Hamiltonian into a dominant contribution and a weaker interaction term,
\begin{equation}
H=H_0+\alpha H_1,
\end{equation}
where $H_0$ represents the exactly solvable part of the Hamiltonian,
$H_1$ contains the perturbative interactions, and
$\alpha\ll1$ characterizes the relative interaction strength.
Unlike conventional product-formula methods, THRIFT treats the evolution
generated by $H_0$ exactly and approximates only the interaction Hamiltonian.
\subsection{Interaction-Picture Transformation}
The quantum state in the interaction picture is defined as
\begin{equation}
|\Psi_I(t)\rangle
=
e^{\,iH_0t}
|\Psi(t)\rangle .
\end{equation}
Substituting this transformation into the Schrödinger equation,
\[
i\frac{d}{dt}|\Psi(t)\rangle
=
(H_0+\alpha H_1)|\Psi(t)\rangle,
\]gives
\begin{equation}
i\frac{d}{dt}
|\Psi_I(t)\rangle
=
H_I(t)
|\Psi_I(t)\rangle,
\end{equation}
where the interaction-picture Hamiltonian is
\begin{equation}
H_I(t)
=
\alpha
e^{\,iH_0t}
H_1
e^{-iH_0t}.
\end{equation}
The formal solution is therefore
\begin{equation}
|\Psi(t)\rangle
=
e^{-iH_0t}
\,
\mathcal{T}
\exp
\left[
-i
\int_0^t
H_I(\tau)
\,d\tau
\right]
|\Psi(0)\rangle,
\end{equation}
where $\mathcal{T}$ denotes the time-ordering operator.
\subsection{THRIFT Approximation}
The time interval is divided into $r$ equal steps,
$\Delta t=t/r$.
Within each interval, the interaction-picture Hamiltonian is assumed to vary
slowly and is approximated by its value at the beginning of the interval,
\begin{equation}
H_I(t)
\approx
H_I(t_j),
\qquad
t_j=j\Delta t.
\end{equation}
The evolution operator for one time step becomes
\begin{equation}
U_{\mathrm{THRIFT}}
(\Delta t)
=
e^{-iH_0\Delta t}
e^{-iH_I(t_j)\Delta t},
\end{equation}
and the complete evolution is
\begin{equation}
U_{\mathrm{THRIFT}}(t)
=
\prod_{j=0}^{r-1}
e^{-iH_0\Delta t}
e^{-iH_I(t_j)\Delta t}.
\end{equation}
\subsection{Error Scaling of THRIFT}
Consider the Hamiltonian
\begin{equation}
H = H_0+\alpha H_1,
\end{equation}
where $H_0$ is the dominant Hamiltonian and
$\alpha H_1$ is a comparatively weak interaction term with
$\alpha\ll1$.
Transforming into the interaction picture gives
\begin{equation}
i\frac{d}{dt}|\Psi_I(t)\rangle
=
H_I(t)|\Psi_I(t)\rangle,
\end{equation}
where
\begin{equation}
H_I(t)
=
\alpha
e^{iH_0t}
H_1
e^{-iH_0t}.
\end{equation}
Since the unitary transformation preserves operator norms,
\begin{equation}
\|H_I(t)\|
=
\alpha\|H_1\|,
\end{equation}
showing that the interaction-picture Hamiltonian remains of order
$\alpha$.
The exact interaction-picture propagator is
\begin{equation}
U_I(t)
=
\mathcal{T}
\exp
\left[
-i
\int_0^t
H_I(\tau)d\tau
\right].
\end{equation}
Using the Magnus expansion,
\begin{equation}
U_I(t)
=
\exp
\left(
\Omega_1+\Omega_2+\Omega_3+\cdots
\right),
\end{equation}
where
\begin{equation}
\Omega_1
=
-i
\int_0^t
H_I(\tau)d\tau
\end{equation}
and
\begin{equation}
\Omega_2
=
-\frac12
\int_0^t
dt_1
\int_0^{t_1}
dt_2
\,
[H_I(t_1),H_I(t_2)].
\end{equation}
The first Magnus term satisfies
\[
\Omega_1
=
\mathcal{O}(\alpha t),
\]
because $H_I=\mathcal{O}(\alpha)$.
For the second term,
\[
[H_I(t_1),H_I(t_2)]
=
\mathcal{O}(\alpha^2),
\]
since each Hamiltonian contributes one factor of $\alpha$.Therefore,
\[
\Omega_2
=
\mathcal{O}
(\alpha^2t^2).
\]
THRIFT integrates the first Magnus term exactly and approximates only the higher-order contributions. Consequently, the leading local truncation error is determined by $\Omega_2$,
\begin{equation}
\epsilon_{\rm local}
=
\mathcal{O}
(\alpha^2\Delta t^2).
\end{equation}
After dividing the total evolution time into
$r$ equal steps,
$\Delta t=t/r$,
the accumulated global error becomes
\begin{equation}
\epsilon_1^{\rm THRIFT}
=
\mathcal{O}
\left(
\frac{\alpha^2t^2}{r}
\right).
\end{equation}
Using symmetric compositions in the interaction picture cancels the
leading odd-order Magnus terms, yielding
\begin{equation}
\epsilon_2^{\rm THRIFT}
=
\mathcal{O}
\left(
\frac{\alpha^2t^3}{r^2}
\right),
\end{equation}
and
\begin{equation}
\epsilon_4^{\rm THRIFT}
=
\mathcal{O}
\left(
\frac{\alpha^2t^5}{r^4}
\right).
\end{equation}
Thus, compared with conventional Trotter--Suzuki formulas, the leading error contains an additional factor of $\alpha$. When the interaction Hamiltonian is weak ($\alpha\ll1$), this substantially suppresses the overall simulation error while preserving the same asymptotic dependence on the number of product-formula steps.
\section{Quantum Circuit Construction and Measurement Procedure}
\label{app:circuit}
The digital quantum simulation is implemented by encoding the discretized
Hermitian Hamiltonian into a register of qubits and approximating its
time-evolution operator using product-formula decompositions. This appendix
describes the mapping of the Hamiltonian to Pauli operators, the construction
of the Trotter and THRIFT circuit and the extraction of computational-basis
probabilities from quantum measurements.
\subsection{Pauli Decomposition of the Hamiltonian}
After discretization of the velocity space and embedding of the resulting
Hermitian Hamiltonian into an $n$-qubit Hilbert space, the Hamiltonian acts on
a Hilbert space of dimension $2^n$. Any Hermitian $n$-qubit Hamiltonian can be
expanded in the Pauli basis as
\begin{equation}
H
=
\sum_{\mu=0}^{4^n-1}
\beta_\mu P_\mu,
\label{eq:pauli_decomposition}
\end{equation}
where
\begin{equation}
P_\mu
\in
\{I,X,Y,Z\}^{\otimes n},
\end{equation}
and the coefficients $\beta_\mu\in\mathbb{R}$ are determined by
\begin{equation}
\beta_\mu
=
\frac{1}{2^n}
\operatorname{Tr}
\left(P_\mu H\right).
\label{eq:pauli_coefficients}
\end{equation}
For the numerical simulations, only the nonzero Pauli coefficients are
retained, giving
\begin{equation}
H
=
\sum_{\mu=1}^{N_P}
\beta_\mu P_\mu,
\label{eq:sparse_pauli}
\end{equation}
where $N_P$ is the number of nonzero Pauli strings. This representation is
particularly convenient for digital quantum simulation because the
evolution generated by an individual Pauli operator can be implemented as a
unitary Pauli rotation.\\
For a single Pauli string $P_\mu$, the corresponding evolution operator is
\begin{equation}
U_\mu(\Delta t)
=
\exp\left(-i\beta_\mu P_\mu\Delta t\right).
\label{eq:pauli_evolution}
\end{equation}
Since
\begin{equation}
P_\mu^2=I,
\end{equation}
this operator can be implemented as a standard Pauli-evolution circuit.
In the Qiskit implementation, Eq.~(\ref{eq:pauli_evolution}) is realized
using the \texttt{PauliEvolutionGate}. Thus, the exponential of each Pauli
term is converted into a native quantum circuit through the corresponding
basis rotations and entangling operations.
\subsection{Quantum Circuit Implementation}
\begin{figure}[h]
    \centering
    \includegraphics[width=\linewidth]{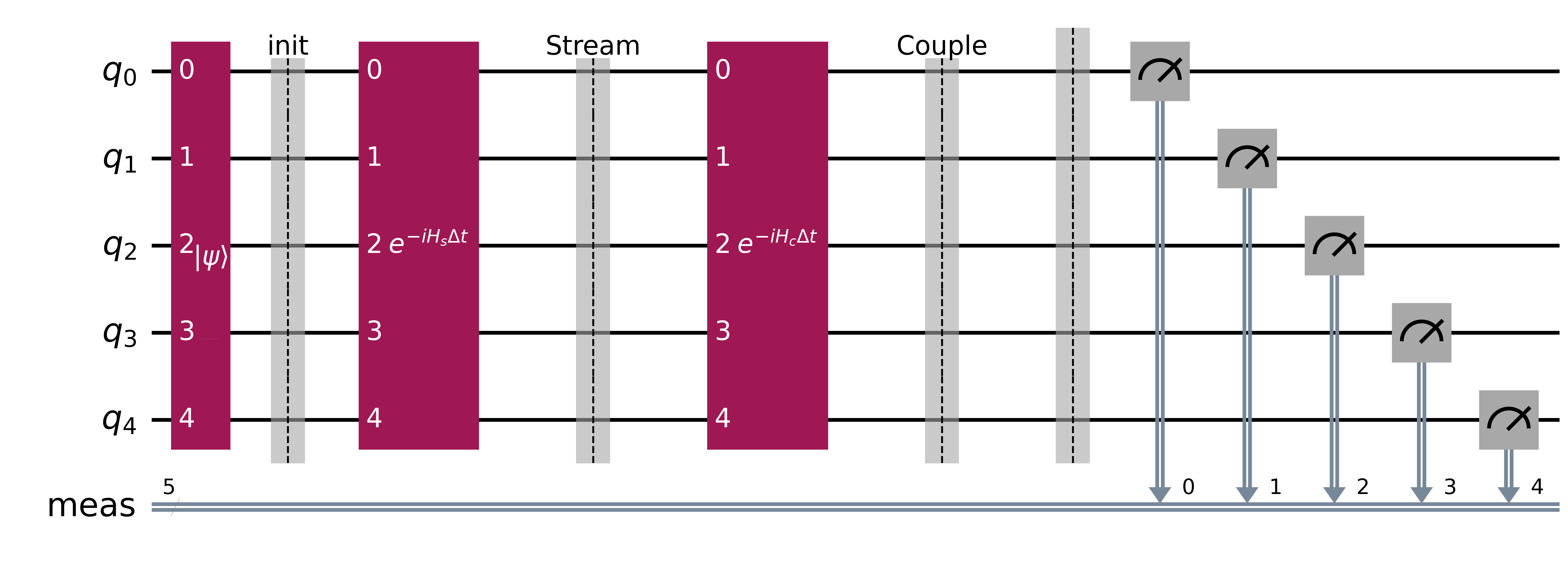}
    \caption{
Representative quantum circuit implementing one first-order Trotter time-evolution step ($r=1$) for the discretized Vlasov--Poisson Hamiltonian. The circuit consists of state preparation, streaming evolution, interaction (coupling) evolution and computational-basis measurement. The implementation shown corresponds to a five-qubit encoding of the discretized Hamiltonian.
}
    \label{fig:appendix_circuit}
\end{figure}

The complete simulation circuit consists of three stages:
(i) preparation of the initial state,
(ii) implementation of the selected product-formula evolution, and
(iii) computational-basis measurement. A representative five-qubit
first-order Trotter circuit is shown in Fig.~\ref{fig:appendix_circuit}.\\
The initial state is prepared as
\begin{equation}
|\Psi(0)\rangle
=
\sum_{j=0}^{2^n-1}
c_j(0)|j\rangle.
\end{equation}
The product-formula circuit then applies the sequence of Pauli-evolution
operators corresponding to the chosen approximation order. For example, a
first-order Trotter step has the form
\begin{equation}
|\Psi_{m+1}\rangle
=
\left[
\prod_{\mu}
e^{-i\beta_\mu P_\mu\Delta t}
\right]
|\Psi_m\rangle,
\end{equation}
and the operation is repeated for $m=0,\ldots,r-1$.\\
The same circuit architecture is used for the second- and fourth-order
Trotter formulas and for the corresponding THRIFT constructions, with only
the sequence of Hamiltonian evolution operators modified according to the
selected product formula.
Figure~\ref{fig:appendix_circuit} illustrates the quantum circuit corresponding to a first-order Trotter decomposition for a single evolution step.
\subsection{Computational-Basis Measurement and Shot Statistics}
Following the unitary evolution, the quantum register is measured in the
computational basis. The final state can be expressed as
\begin{equation}
|\Psi(t)\rangle
=
\sum_{j=0}^{2^n-1}
c_j(t)|j\rangle .
\label{eq:final_state}
\end{equation}
The ideal probability of obtaining the computational-basis state $|j\rangle$
is given by Born's rule,
\begin{equation}
p_j
=
|\langle j|\Psi(t)\rangle|^2
=
|c_j(t)|^2,
\label{eq:born_probability}
\end{equation}
with
\begin{equation}
\sum_{j=0}^{2^n-1}p_j=1.
\end{equation}
On a quantum processor, these probabilities are estimated from repeated
measurements or shots. For a total of $N_{\mathrm{shots}}$ measurements, if
the state $|j\rangle$ is observed $N_j$ times, its estimated probability is
\begin{equation}
\widehat{p}_j
=
\frac{N_j}{N_{\mathrm{shots}}}.
\label{eq:shot_probability}
\end{equation}
The statistical uncertainty associated with the estimated probability is,to leading order,
\begin{equation}
\sigma_j
\simeq
\sqrt{
\frac{p_j(1-p_j)}
{N_{\mathrm{shots}}}
}.
\label{eq:shot_error}
\end{equation}
Thus, increasing the number of shots reduces statistical sampling
uncertainty as $N_{\mathrm{shots}}^{-1/2}$.\\
For example, the probability of the basis state $|11111\rangle$ is estimated
from
\begin{equation}
\widehat{p}_{11111}
=
\frac{N_{11111}}{N_{\mathrm{shots}}}.
\label{eq:p11111}
\end{equation}
The corresponding numerical results for the different product-formula
methods are compared in Fig.~\ref{fig:appendix_basis}.
\begin{figure}[h]
    \centering
    \includegraphics[width=\linewidth]{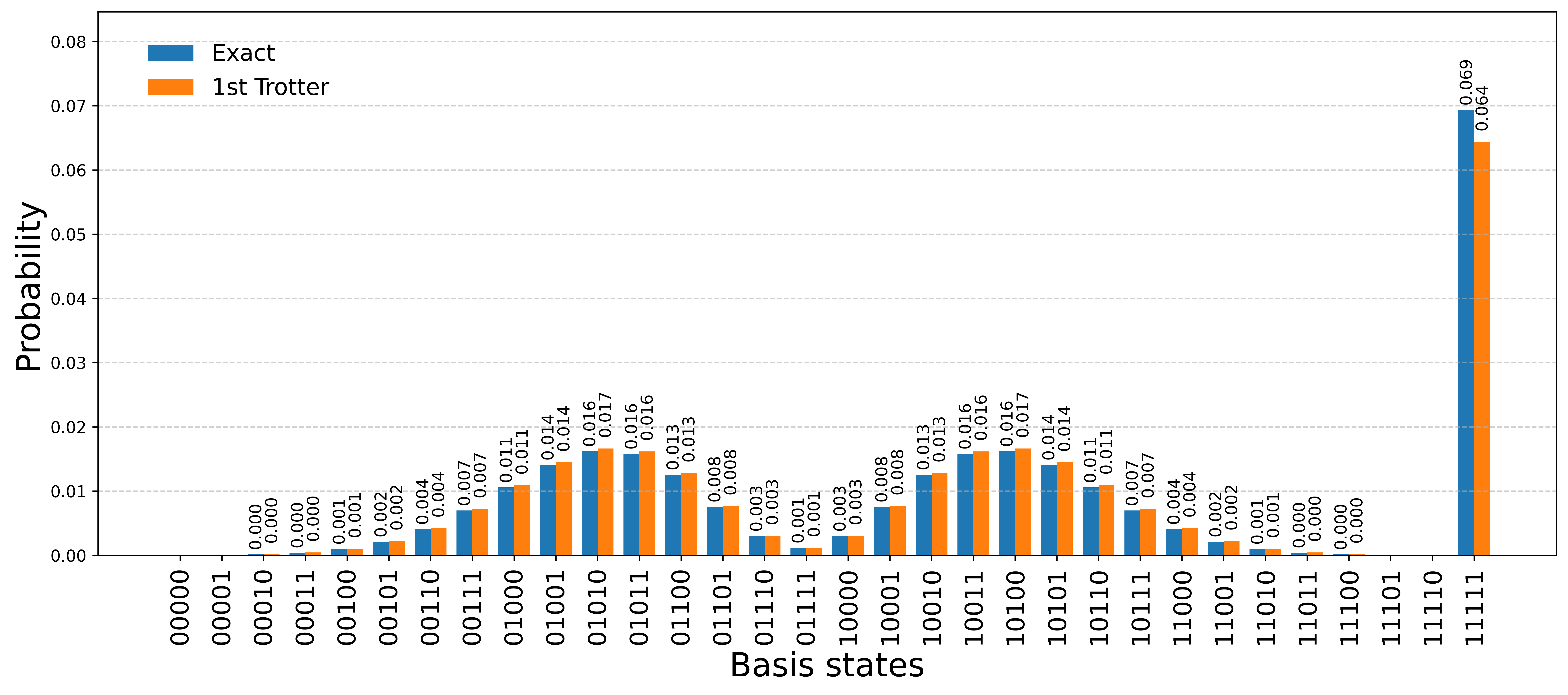}
    \caption{
Probability distribution over the computational basis states after quantum evolution. The blue bars denote the exact Hamiltonian evolution, while the orange bars correspond to the first-order Trotter approximation. The close agreement confirms that the product-formula circuit accurately reproduces the evolved quantum state.
}
    \label{fig:appendix_histogram}
\end{figure}
Figure~\ref{fig:appendix_histogram} compares the probability distribution obtained from the exact Hamiltonian evolution with that produced by the first-order Trotter approximation. The excellent agreement between the two distributions demonstrates that the quantum circuit accurately reproduces the exact state populations. Small deviations arise from the finite product-formula approximation and are consistent with the theoretical Trotter error discussed in Appendix~\ref{app:trotter}.
\subsection{Statevector and Shot-Based Evaluation}

Two complementary approaches are used in the numerical analysis. In
statevector simulations, the complete complex state
$|\Psi(t)\rangle$ is available and the exact computational-basis probabilities
are obtained directly from Eq.~(\ref{eq:born_probability}). This representation
is used to quantify the intrinsic product-formula error without statistical
sampling noise.\\
In contrast, shot-based simulations reproduce the measurement procedure of a
quantum processor. Only the measurement outcomes are available and the
probabilities are reconstructed according to Eq.~(\ref{eq:shot_probability}).
Consequently, shot-based results contain finite-sampling fluctuations in
addition to the deterministic product-formula error.
The distinction is important when evaluating complex amplitudes. A computational-basis measurement directly provides $|c_j|^2$, but does not provide the complex phase of $c_j$. Therefore, quantities that depend on the
complex amplitudes themselves cannot, in general, be reconstructed from a single computational-basis measurement. In the present classical statevector simulations, the complex amplitudes are directly accessible and
are therefore used for evaluating amplitude-dependent observables and
state-vector errors.
\subsection{Extraction of the Electric-Field Observable}
The electric-field degrees of freedom are represented by designated
components of the encoded state. For the statevector simulation, if the
components associated with the electric-field variables are denoted by
$c_{E_x}(t)$ and $c_{E_y}(t)$, their amplitudes are directly obtained from
the simulated state,
\begin{equation}
E_x(t)=c_{E_x}(t),
\qquad
E_y(t)=c_{E_y}(t).
\label{eq:electric_amplitudes}
\end{equation}
The corresponding electric-field magnitude is then
\begin{equation}
|E(t)|^2
=
|E_x(t)|^2+|E_y(t)|^2.
\label{eq:electric_energy}
\end{equation}
\begin{figure}[h]
    \centering
    \includegraphics[width=\linewidth]{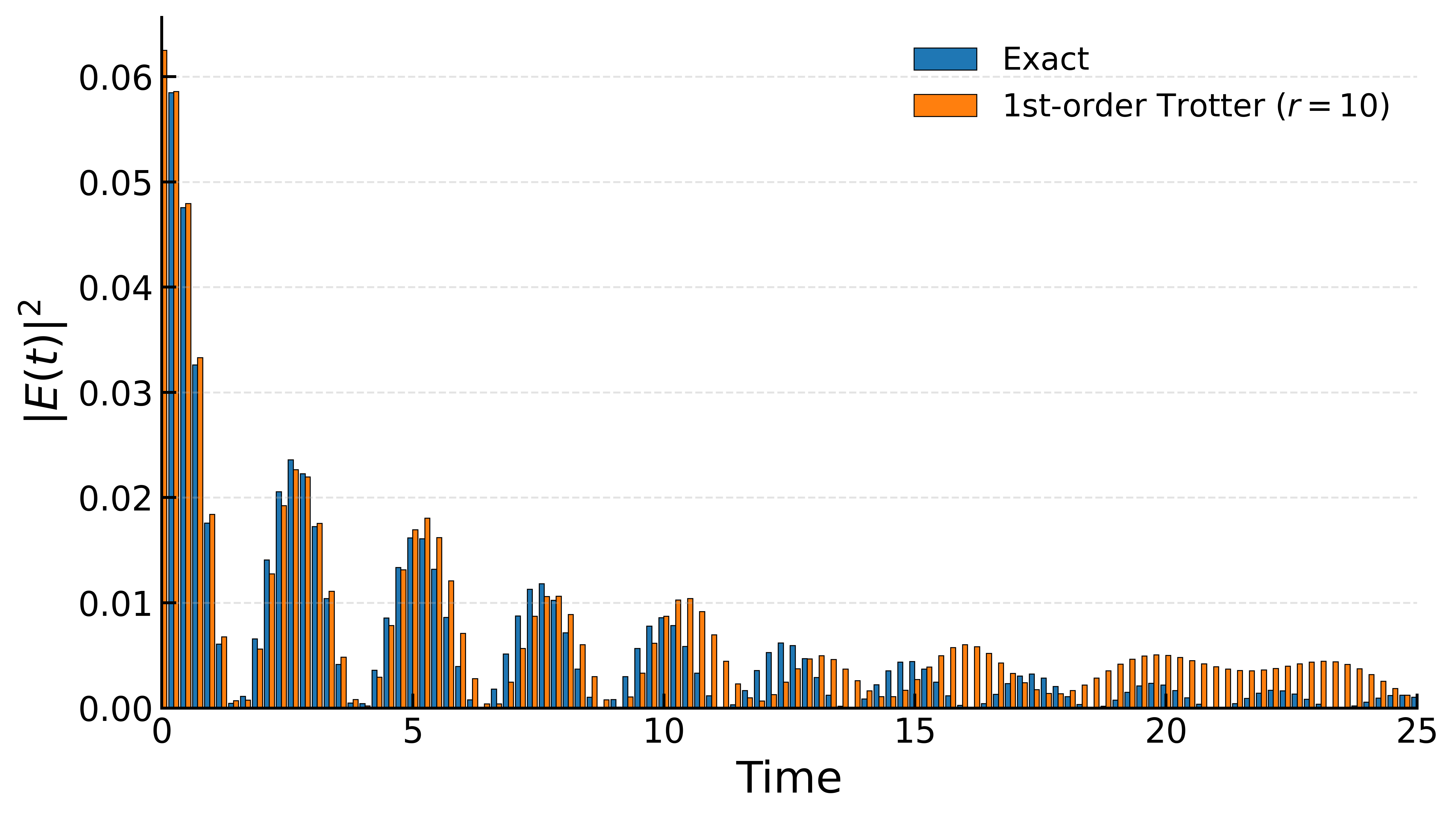}
    \caption{
Comparison of the electric-field energy extracted from the exact Hamiltonian evolution and the first-order Trotter simulation. The observable is reconstructed from the amplitudes of the electric-field basis states encoded in the quantum register.
}
    \label{fig:appendix_energy}
\end{figure}
The comparison between the exact evolution and the first-order Trotter
simulation is shown in Fig.~\ref{fig:appendix_energy}. Agreement between the two results provides a direct test of the ability of the product-formula
circuit to reproduce the relevant physical observable.\\
For an actual quantum-hardware implementation, expectation values of
observables are obtained from repeated measurements of the corresponding
Hermitian operators. For an observable $O$, the expectation value is
\begin{equation}
\langle O\rangle_t
=
\langle\Psi(t)|O|\Psi(t)\rangle,
\end{equation}
and, when $O$ is decomposed into Pauli operators,
\begin{equation}
O=\sum_{\nu}a_\nu Q_\nu,
\end{equation}
its expectation value can be reconstructed from measurements of the
individual Pauli operators,
\begin{equation}
\langle O\rangle_t
=
\sum_{\nu}a_\nu
\langle Q_\nu\rangle_t.
\label{eq:observable_pauli}
\end{equation}
\subsection{Consistency of the Measurement Procedure}
The probability distributions obtained from the exact evolution and the first-order Trotter circuit are compared in Fig.~\ref{fig:appendix_histogram}. The close correspondence between the two
distributions demonstrates that the product-formula approximation preserves
the dominant computational-basis populations for the parameters considered.
\begin{figure}[h]
    \centering
    \includegraphics[width=\linewidth]{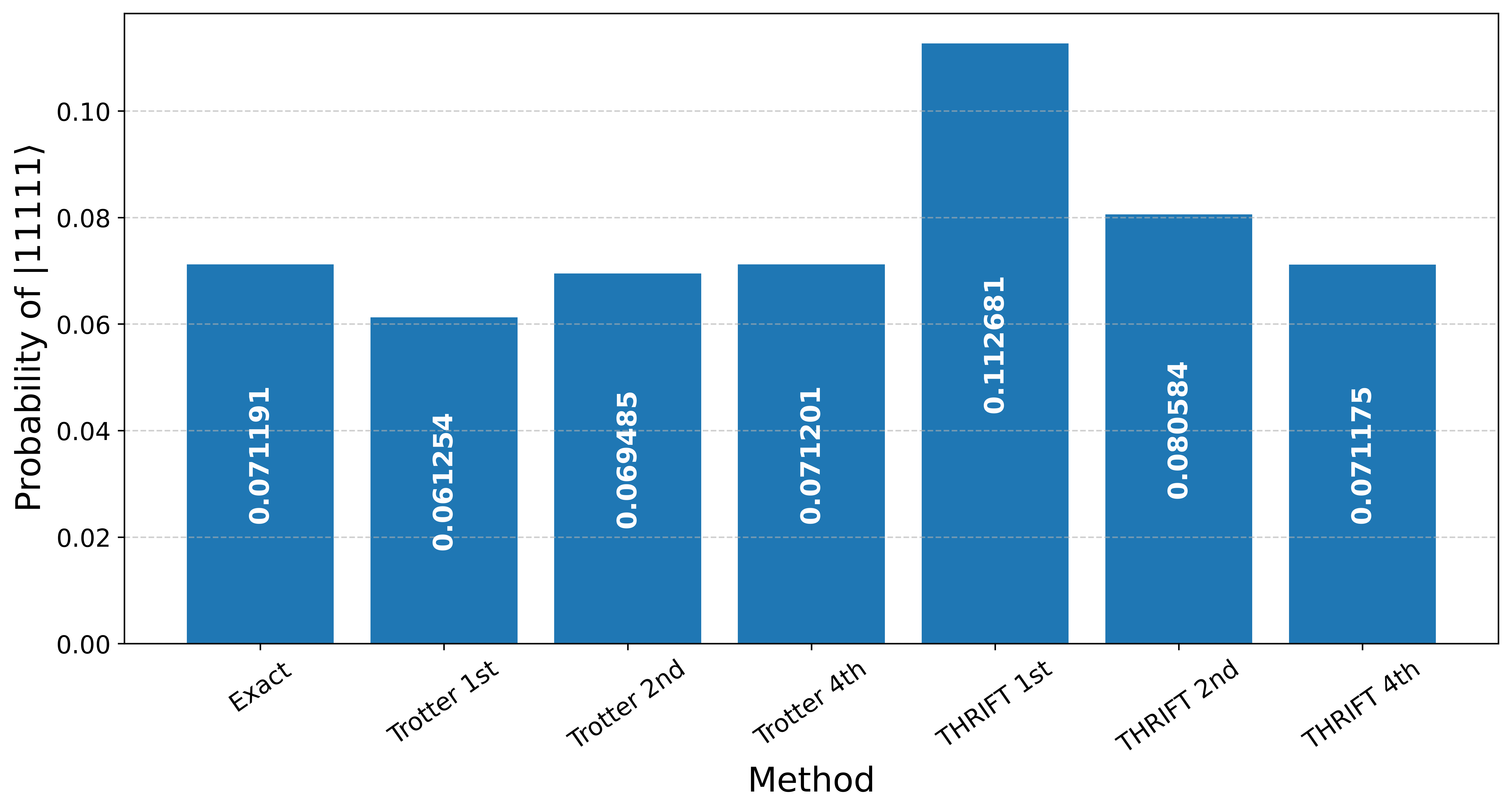}
    \caption{
Probability of the dominant computational basis state $|11111\rangle$ obtained from the exact Hamiltonian evolution together with first, second and fourth-order Trotter and THRIFT simulations. Higher-order product-formula methods reproduce the exact population more accurately than first-order approximations.
}
    \label{fig:appendix_basis}
\end{figure}
The same measurement procedure is applied to all product-formula approximations shown in \ref{fig:appendix_basis}. For each method, the Hamiltonian is first mapped to its
corresponding Pauli representation, the required product-formula sequence is
constructed from Pauli-evolution operators, and the resulting circuit is
evolved for the selected number of steps $r$. Computational-basis
measurements are then performed over a prescribed number of shots, from
which the probability distribution is reconstructed using
Eq.~(\ref{eq:shot_probability}). This common procedure allows the Trotter and
THRIFT algorithms to be compared on the same basis in terms of simulation
accuracy and quantum resources.
\section{Richardson Extrapolation for Trotter and THRIFT Error Reduction}
\label{app:richardson}
Richardson extrapolation (RE) provides a systematic procedure for suppressing the leading discretization error of product-formula approximations through classical post-processing. Let an approximation
obtained with time step $h$ admit the asymptotic expansion
\begin{equation}
A(h)
=
A
+
c_p h^p
+
\mathcal{O}(h^{p+1}),
\label{eq:rich_expansion}
\end{equation}
where $A$ is the exact result, $p$ is the order of the underlying
product formula, and $c_p$ is independent of $h$. Evaluating the
approximation at $h$ and $h/2$ gives
\begin{equation}
A(h/2)
=
A
+
c_p\left(\frac{h}{2}\right)^p
+
\mathcal{O}(h^{p+1}).
\end{equation}
The leading order error is therefore eliminated by the linear combination
\begin{equation}
A_{\mathrm{RE}}
=
\frac{2^p A(h/2)-A(h)}
{2^p-1},
\label{eq:richardson}
\end{equation}
yielding
\begin{equation}
A_{\mathrm{RE}}
=
A+\mathcal{O}(h^{p+1}).
\label{eq:rich_error}
\end{equation}
For the Hamiltonian simulations considered here, the time step is
\begin{equation}
h=\frac{t}{r},
\end{equation}
where $t$ is the total evolution time and $r$ is the number of
product-formula steps. Consequently, increasing $r$ from $r$ to $2r$
corresponds to $h\rightarrow h/2$. If
$|\psi_r\rangle$ denotes the state obtained with $r$ product-formula
steps, the Richardson-extrapolated state is defined as
\begin{equation}
|\psi_{\mathrm{RE}}\rangle
=
\frac{
2^p|\psi_{2r}\rangle-|\psi_r\rangle
}{
2^p-1
}.
\label{eq:richardson_state}
\end{equation}
Because the linear combination in Eq.~(\ref{eq:richardson_state}) is
not, in general, normalized, the resulting state is normalized before
the evaluation of state-dependent quantities,
\begin{equation}
|\widetilde{\psi}_{\mathrm{RE}}\rangle
=
\frac{|\psi_{\mathrm{RE}}\rangle}
{\sqrt{\langle\psi_{\mathrm{RE}}|\psi_{\mathrm{RE}}\rangle}}.
\label{eq:richardson_normalization}
\end{equation}
For the first, second and fourth-order product formulas employed in this work, Eq.~(\ref{eq:richardson_state}) respectively becomes
\begin{align}
|\psi_{\mathrm{RE}}^{(1)}\rangle
&=
2|\psi_{2r}\rangle-|\psi_r\rangle,
\\[2mm]
|\psi_{\mathrm{RE}}^{(2)}\rangle
&=
\frac{4|\psi_{2r}\rangle-|\psi_r\rangle}{3},
\\[2mm]
|\psi_{\mathrm{RE}}^{(4)}\rangle
&=
\frac{16|\psi_{2r}\rangle-|\psi_r\rangle}{15}.
\end{align}
The same extrapolation procedure is applied to both the Trotter and THRIFT product-formula simulations. For a fixed total evolution time,
the extrapolation combines two simulations with different temporal resolutions and does not modify the circuit depth of either individual
circuit. Its effect is instead realized through classical post-processing of the corresponding simulation results. The leading
product-formula contribution is thereby cancelled, improving the convergence with respect to the number of evolution steps without
altering the underlying quantum circuit construction.

\end{document}